\documentclass{tADP2e}
\usepackage{textcomp}
\usepackage{amsmath}
\usepackage{amssymb}
\usepackage{graphicx}

\begin{document}

\title{Decoherence in Solid State Qubits}
\author{Luca Chirolli and Guido Burkard
\footnote{chirolli@physik.rwth-aachen.de, burkard@physik.rwth-aachen.de}\\
Department of Physics, University of Konstanz, D-78457
  Konstanz, Germany\\
Institute of Theoretical Physics C, RWTH Aachen, 
D-52056 Aachen, Germany}

\maketitle

\begin{abstract}
Interaction of solid state qubits with environmental degrees of
freedom strongly affects the qubit dynamics, and leads to
decoherence. In quantum information processing with solid state
qubits, decoherence significantly limits the performances of such 
devices. Therefore, it is necessary to fully understand the mechanisms
that lead to decoherence. In this review we discuss how decoherence
affects two of the most successful realizations of solid state qubits,
namely, spin-qubits and superconducting qubits. In the former,
the qubit is encoded in the spin 1/2 of the electron, and it is
implemented by confining the electron spin in a semiconductor quantum
dot. Superconducting devices show quantum behavior at low
temperatures, and the qubit is encoded in the two lowest energy levels
of a superconducting circuit. The electron spin in a quantum dot has
two main decoherence channels, a (Markovian) phonon-assisted
relaxation channel, due to the presence of a spin-orbit interaction,
and a (non-Markovian) spin bath constituted by the spins of the nuclei
in the quantum dot that interact with the electron spin via the
hyperfine interaction. In a superconducting qubit, decoherence takes
place as a result of fluctuations in the control parameters, such as
bias currents, applied flux, and bias voltages, and via losses in the
dissipative circuit elements.     
\end{abstract}


\section{Introduction}

\subsection{What is coherence? and why is it interesting?}

Coherence is a defining property of quantum mechanics. 
It can be argued that quantum coherence is the property that
draws a line between the ``quantum world'' from the ``classical
world''. But what exactly is coherence?   
In physics, the term coherence refers to the property of waves to
interfere, showing well known interference patterns. Two waves,
depending on their relative phase, can produce a constructive
interference, characterized by an enhancement of the amplitude of the
wave, or destructive interference,  accompanied by a complete
suppression of it. Only the relative phase of the two waves makes the
difference. To be precise we should therefore speak about phase
coherence of quantum states.  

By quantum states here we mean states of a quantum system, which in
turn can be constituted by more than one quantum object. The same
rules of quantum mechanics that allow us to explain and predict
interference of one object with itself, as the case of an electron
through a double slit, predict that a system composed by two quantum
subsystems can be in a state that has no classical counterpart, being
a superposition with a precise phase of two or more quantum
states of the whole system. This property of quantum states goes under
the name of entanglement, and entangled states need to be phase
coherent.  
In particular, coherence, as a property of quantum
mechanical phenomena, disappears in the classical world, and it is 
therefore of fundamental interest to study it in theory and
experiment. 
It is never completely possible to isolate a quantum system from the
surrounding world. The system and its surrounding interact and, as a
result, a randomization of the phase of the quantum system takes
place, and the initial quantum state ends up in a classical
state. This process is known as decoherence. 

Only in recent years, thanks to the advances in technology, it has
become possible to study quantum effects involving single
quantum objects, like single photons, ions, electron spins,
etc... Particular attention has been devoted to see coherence, from an
experimental point of view, and to understand its limitation. In fact,
though remarkable improvements have been achieved, nowadays, to see
coherence, a lot of effort must be spent to understand how to
preserve coherence. 
In the last decades, the idea of joining quantum physics laws and
information science gave birth to a new and intriguing branch of
science, quantum information theory, which studies the possibilities
that  quantum rules offer to information processing. In particular, the
superposition principle opens the possibility to perform new and fast
algorithms. 
The physical implementation of quantum information processing
represents a challenge because one has to deal with the competition
between fast and reliable quantum control, that requires interaction
with the outside world, and good isolation of the quantum devices in
order to ensure long coherence times.

Therefore, it is important to understand theoretically
how decoherence happens in the systems under study (here, solid state
systems) in order to make progress toward this
ambitious goal (i.e. implementing quantum information).

\subsubsection{The quantum bit}

Classical information is based on binary logic, in which information
is encoded in a series of bits (binary digits) that can assume only
two values, 0 or 1. A typical example is a switch, with its two
possible states ``on'' and ``off''. All classical logical operations
can be implemented as algorithms based on one- and two-bit operations,
the so-called gates.    

The building block of quantum information is the quantum bit,
or qubit. Using the Dirac notation, the two states that characterize the
qubit are $|0\rangle$ and $|1\rangle$ and they represent the quantum
counterpart of the classical 0 and 1. The most important property of a
quantum bit is the possibility to be in a coherent superposition
state 
\begin{equation}
|\psi\rangle=\alpha|0\rangle+\beta|1\rangle,
\end{equation}
with $\alpha$ and $\beta$ complex numbers, characterized by a relative
phase and by $|\alpha|^2+|\beta|^2=1$. According the the postulates of
quantum mechanics, $|\alpha|^2$ represents the probability for the
qubit to be in the state $|0\rangle$, whereas  $|\beta|^2$ represents
the probability to be in $|1\rangle$. This means that if we prepare
many copies of the same system in the state $|\psi\rangle$, a
measurement of the state of the qubit will produce the outcome 0 with
rate $|\alpha|^2$, and the outcome 1 with rate $|\beta|^2$. In which
case the two logical states are the spin up $|\uparrow\rangle$ and the
spin down $|\downarrow\rangle$. 
The two states $|0\rangle$ and $|1\rangle$ form a basis of the Hilbert
space ${\cal H}={\rm span}\{|0\rangle,|1\rangle\}$ of the qubit. 

A good example of a qubit is the spin 1/2.
In order to explain the necessity to use complex number $\alpha$ and
$\beta$ to characterize the state of the qubit, we describe an
interference procedure for a spin 1/2 particle. Suppose we prepare the
spin in the state $|\psi_0\rangle=|\uparrow\rangle$, that is with
probability 1 to find it parallel with respect to a certain direction
$z$ in the space, that we choose as quantization axis. We then rotate
the spin by an angle $\pi/2$ about an axis perpendicular to $z$,
i.e. the $y$ axis. The result is the state 
\begin{equation}
|\psi_1\rangle=e^{-i\frac{\pi}{4}\sigma_y}|\uparrow\rangle=
\frac{1}{\sqrt{2}}(|\uparrow\rangle+|\downarrow\rangle).
\end{equation} 
We then let the spin cross a region in which there is a magnetic field
that points in the positive $z$ direction, ${\bf B}=(0,0,B)$. Due to
the presence of the magnetic field, the two states $|\uparrow\rangle$
and $|\downarrow\rangle$ accumulate a relative phase $2\varphi$, that
depends on the magnitude of the magnetic field and the time $t$ spent
in the region with the ${\bf B}$ field, and that for simplicity we
leave unspecified. Up to an overall phase, the state of the system
that comes out from the region with a magnetic field is given by 
\begin{equation}
|\psi_2\rangle=\frac{1}{\sqrt{2}}\left(|\uparrow\rangle+
e^{2i\varphi}|\downarrow\rangle\right).
\end{equation} 
Now, we again rotate the spin of $\pi/2$ about the $y$ direction,
and obtain  
\begin{equation}
|\psi_3\rangle=e^{-i\frac{\pi}{4}\sigma_y}|\psi_2\rangle=
e^{i\varphi}\left[\cos(\varphi)|\uparrow\rangle
+i\sin(\varphi)|\downarrow\rangle\right].
\end{equation}
If we now measure the state of the spin, we obtain $|\uparrow\rangle$
with probability $\cos^2(\varphi)$ and $|\downarrow\rangle$
with probability  $\sin^2(\varphi)$. We clearly see, now, that the
relative phase can really affect the state of a quantum system. 
This procedure is known as Ramsey interference \cite{Ramsey-PR50} and
it is used in experiments to detect coherent oscillations in the
transverse spin component.

\subsubsection{One qubit as environment}

Decoherence is a consequence of the interaction of the qubit with the
surrounding environment. As an instructive example we consider the case
in which the environment is constituted by another qubit. For the
Hamiltonian describing the interaction between the two qubits we
choose ($\hbar=1$)
\begin{equation}\label{Eq:sYsY}
H=\frac{J}{4}\sigma^z_1\otimes\sigma^z_2,
\end{equation}
where the operator $\sigma^z_1\otimes\sigma^z_2$ is a two-qubit operator,
given by the tensor product of two single-qubit operators, and it acts
in the tensor product space ${\cal H}={\cal H}_1\otimes{\cal H}_2$. We
note that our general argument does not depend on the specific form of
$H$, as long as it describes an interaction between the qubits.
As the initial state for the two qubit system we choose a product
state $|+\rangle_1\otimes|+\rangle_2$, where the single qubit state is 
$|+\rangle=(|0\rangle+|1\rangle)/\sqrt{2}$, written in the basis
diagonal with respect to $\sigma_z$, $\sigma_z|0\rangle=|0\rangle$,
and  $\sigma_z|1\rangle=-|1\rangle$. 
We let the system evolve according to the unitary evolution generated
by the Hamiltonian Eq.~(\ref{Eq:sYsY}) for a time $t$, after which we
perform a trace operation on the second qubit and have a look how the
state of the first qubit evolved during the time $t$ in which it has
interacted with the second qubit. We re-write the initial state of the
first qubit as a pure state density matrix, $\rho_1=|+\rangle_1\langle
+|$. In the $\{|0\rangle,|1\rangle\}$ basis it is found to be 
\begin{equation}
\rho_1(0)=\frac{1}{2}
\left(\begin{array}{cc}
1 & 1\\
1 & 1\end{array}\right).
\end{equation}
The state of the two qubit system after a time $t$ is given by
$|\psi(t)\rangle=U(t)|++\rangle$, with
$U(t)=\exp(-iJt\sigma^z_1\sigma^z_2/2)$. After some algebra the state
of the first qubit at time $t$ is given by
\begin{equation}
\rho_1(t)={\rm Tr}_2[|\psi(t)\rangle\langle\psi(t)|]=
\frac{1}{2}\left(\begin{array}{cc}
1 & \cos(Jt/2)\\
\cos(Jt/2) & 1\end{array}
\right).
\end{equation}  
The diagonal element of the first qubit density matrix is left
unchanged by the interaction with the second qubit, whereas the
off-diagonal elements change in time. The coherence of a state is
encoded in the off-diagonal element of the density matrix. After a
time $t=\pi/J$ the coherence is completely lost (full
decoherence). However, due to the smallness of the environment
considered, the first qubit periodically recovers its original
state. It is therefore clear that the interaction with the environment
strongly affects the qubit coherence.

\subsection{Quantum open systems} 

According to the axioms of quantum mechanics the dynamics of a closed
conservative system is described as a unitary time evolution. In such
a picture the system is considered to be decoupled from the
surrounding environment, that does not influence at all the dynamics
of the closed system. Strictly speaking this is never the
case. However, under certain conditions the coupling to the
environment can be considered to be weak, and to a good approximation
neglected. 

In condensed phases, the coupling to the environment can be relatively
strong, and the system under consideration cannot be separated from
its surrounding. However, often a rather complex physical situation
can be modelled by a system that consists of few dynamical variables
in contact with a huge environment, constituted by a very large or
even infinite number of degrees of freedom. In this case the small
relevant system alone has to be described as an open system.

Generally, an open system is a
quantum system $S$ which is coupled to an other quantum system $B$
called the environment. It can therefore be thought to be a subsystem
of the combined system $S+B$, which in turn in many cases is considered
to be a closed system, governed by Hamiltonian dynamics. The system $S$
will in turn change according to its internal dynamics, and as a
consequence of the interaction with the environment. Certain
system-environment correlations will be established between the two parts
and, as a consequence, the dynamics of a quantum open system cannot,
in general, be described in terms of a unitary time evolution.

Denote ${\cal H}_S$ the Hilbert space of the system $S$, and ${\cal
  H}_B$ the Hilbert space of system $B$. The dynamics of the combined
system $S+B$ takes place in the Hilbert space given by the tensor
product space ${\cal H}_{SB}={\cal H}_S\otimes{\cal H}_B$. The total
Hamiltonian can be chosen to have the general form 
\begin{equation}
H=H_S+H_B+H_I,
\end{equation}
where $H_S$ describes the evolution of the system $S$ alone, $H_B$ is
the free Hamiltonian of the environment $B$, and $H_I$ describes the
interaction between the system and the environment. Usually when
speaking about the {\it environment} of the system $S$, the term {\it
  reservoir} may appear, that refers to an environment with an infinite
number of degrees of freedom, such that the frequency modes associated
with it form a continuum spectrum. Occasionally the term {\it heat
  bath} or simply {\it bath} refers to a reservoir which is in thermal
equilibrium. 

The presence of an environment is meant to model the communication of
the open system with the external world. However the attention is focused
on the subsystem under study $S$, and all observations of interest
refer to the subsystem $S$. Formally this means that all observables
of interest act on the Hilbert space ${\cal H}_S$. Denoting
the state of the total system by $\rho$, the expectation values
of all observables may be written as
\begin{equation}
\langle{\cal O}\rangle={\rm Tr}_S\left[{\cal O}\rho_S\right],
\qquad \rho_S={\rm Tr}_B[\rho],
\end{equation} 
where ${\cal O}$ is the Hermitian operator describing the observable,
$\rho_S$ is the reduced density matrix of the open system $S$, 
and ${\rm Tr}_{S(B)}$ denotes a partial trace on the system $S(B)$. 

All informations that describe the open system $S$ are contained in
the reduced density matrix $\rho_S$. Since the total system evolves
unitarily in time, $\rho_S(t)$ is obtained as partial trace over the
environment $B$ of $\rho(t)$,
\begin{equation}
\rho_S(t)={\rm Tr}_B\left[U(t,t_0)\rho_S(t_0)U^{\dag}(t,t_0)\right],
\end{equation}
where $U(t,t_0)$ is the unitary evolution operator of the total system.  
The equation of motion for the open system reduced density matrix
$\rho_S(t)$ is
\begin{equation}
\frac{d}{dt}\rho_S(t)=-i{\rm Tr}_B\left[H(t),\rho_S(t)\right].
\end{equation}    

\subsection{Generalized master equation}

In many cases it is useful to model the dynamics of an open system by
means of an appropriate equation of motion for its density matrix, the
so called quantum master equation.  The evolution in time of the
total system $\rho$ is governed by the well known Liouville equation
of motion  
\begin{equation}\label{Eq:Liouville-eq-TotSys}
\dot{\rho}=-i[H(t),\rho(t)]\equiv-i{\cal L}\rho(t),
\end{equation}
where the second equality defines the Liouville operator ${\cal L}$. 
As the Hamiltonian can be divided into three terms that describe the
dynamics of the two systems alone, $H_S$ and $H_B$, and a interaction
between the two parts, $H_I$, the Liouville operator can be written as
the sum of three contributions 
\begin{equation}
{\cal L}={\cal L}_S+{\cal L}_B+{\cal L}_I.
\end{equation}
Without going into details that are beyond the scope of this review,
we just mention the fact that the Liouvillian is a superoperator, that
maps operators into operators. The initial state for the combined
system $S+B$ can typically be chosen to be a product state,
$\rho(0)=\rho_S(0)\otimes\rho_B$.

We have already introduced the reduced density matrix $\rho_S$ of the
open subsystem $S$. It can be formally obtained from the density
matrix of the total system $\rho$ by means of a projection operation,
that contains a partial trace over the system $B$, 
\begin{equation}\label{Eq:initialrho}
\rho_S={\cal P}\rho={\rm Tr}_B[\rho]\otimes\rho_B.
\end{equation}
Here, $\rho_B$ is a fixed density matrix for the environment. 
Mapping operators into operators, the projector ${\cal P}$ is also a
superoperator. We may thus decompose the $\rho$ as 
\begin{equation}
\rho(t)=\rho_S(t)+(1-{\cal P})\rho(t),\qquad {\cal P}^2={\cal P}.
\end{equation}
Substituting this decomposition in the Liouville equation of motion
for the total system Eq.~(\ref{Eq:Liouville-eq-TotSys}), choosing the
projector in such a way that the inhomogeneous term that depends on the
initial state can be disregarded, and using that the operator ${\cal P}$
defined in Eq.~(\ref{Eq:initialrho}) commutes with the Liouvillian of
the open system ${\cal L}_S$, after some 
algebra, the equation of motion for the reduced density matrix
$\rho_S(t)$ can be cast in the form of an exact generalized master
equation, the {\it Nakajima-Zwanzig equation}
\cite{Nakajima-ProgrThPhy58,Zwanzig-JChmPhys60} 
\begin{eqnarray}\label{Eq:Nakajima-Zwanzig-GME}
\dot{\rho}_S(t)&=&-i{\cal L}_S\rho_S(t)+\int_0^tdt'
\Sigma(t-t')\rho_S(t'),\\ 
\Sigma(t)\rho_S&=&-i{\rm Tr}_B\left[{\cal L}_I
e^{(1-{\cal P}){\cal L}t}{\cal L}_I\rho_S\otimes\rho_B\right],
\end{eqnarray}
where $\Sigma(t)$ is the self-energy superoperator. The first term
describes the reversible evolution of open system $S$, while the second
term produces irreversibility.

\subsubsection{Born approximation}

The generalized master equation Eq.~(\ref{Eq:Nakajima-Zwanzig-GME}) 
is a formally exact
and closed description of the dynamics of the state of the system
$\rho_S$, but it is very complicated from a mathematical point of view
and rather unpractical. Usually, in order to handle it and
some approximation are made. In fact, the kernel of
Eq.~(\ref{Eq:Nakajima-Zwanzig-GME}) contains all powers of ${\cal
  L}_I$, and the 
dynamics of $\rho_S$ at time $t$ depends on the whole history of the
density matrix. If the coupling between system and reservoir is weak,
i.e. $\left\|{\cal L}_I\right\|\ll\left\|{\cal L}_S+{\cal L}_B
\right\|$, the exponential can be expanded in power of ${\cal L}_I$
in a perturbative way. In lowest order Born approximation, the
interaction Liouvillian is disregarded in the exponent and 
${\cal L}_I$ is retained only to second order
\begin{equation}\label{Eq:GME-Born-app}
\tilde{\Sigma}(t)\rho_S=-i{\rm Tr}_B\left[{\cal L}_I
e^{(1-{\cal P})({\cal L}_S+{\cal L}_B)t}{\cal L}_I\rho_S
\otimes\rho_B\right].
\end{equation} 
The applicability of the master equation in the Born approximation is
strictly restricted to those cases in which the coupling between
system and environment is weak, with decoherence and relaxation times
large compared 
to the relevant time scales of the reversible dynamics.

\subsection{Quantum Markov process}

The master equation in the Born approximation
Eq.~(\ref{Eq:GME-Born-app}), though it is much simpler than the exact
Nakajima-Zwanzig equation (\ref{Eq:Nakajima-Zwanzig-GME}), is still an
integro-differential equation that is very difficult to
handle. Assuming that 
the temporal correlations in the bath are short lived and typically
lead to exponential decay of the coherence and populations, the master
equation in the Born approximation Eq.~(\ref{Eq:GME-Born-app}) can be
further 
simplified. In the Born-Markov approximation the master equation for
the reduced density matrix of system $S$ assumes the form
\begin{eqnarray}\label{Eq:BornMarkov-ME}
\dot{\rho}_S(t)&=&-i{\cal L}_S\rho_S(t)+\tilde{\Sigma}^R(t)\rho_S(t),\\ 
\tilde{\Sigma}^R(t)&=&-i\int_0^tdt'\tilde{\Sigma}(t')e^{it'{\cal L}_S}.
\end{eqnarray}

In an eigenstate basis of $H_S$, the master equation in the
Born-Markov approximation can be written as the so called Redfield
equation \cite{Redfield-65,Blum81,Slichter}
\begin{equation}\label{Eq:Redfield-eq}
\dot{\rho}_{nm}=-i\omega_{nm}\rho_{nm}(t)-\sum_{k,l}R_{nmkl}\rho_{kl}(t),
\end{equation}
where $\rho_{nm}=\langle n|\rho_S|m\rangle$, 
$\omega_{nm}=\omega_n-\omega_m$, and we have introduced the Redfield
tensor
\begin{equation}
R_{nmkl}=\int_0^{\infty}dt{\rm Tr}_B\left[\langle n|\left[
H^{\rm int}_I(t),\left[H^{\rm int}_I(0),|k(t)\rangle\langle l(t)|
\rho_B\right]\right]|m\rangle\right],
\end{equation}
where we have used the interaction picture Hamiltonian and the system
eigenstates in the interaction picture
\begin{equation}
H^{\rm int}_I(t)=e^{i(H_S+H_B)t}H_Ie^{-i(H_S+H_B)t},
\qquad |k(t)\rangle=e^{iH_St}|k\rangle=e^{i\omega_kt}|k\rangle.
\end{equation}
The first term of Eq.~(\ref{Eq:Redfield-eq}) represents the reversible
motion in terms of the transition frequencies $\omega_{nm}$, while the
second term describes relaxation. The Redfield tensor can be expressed
as
\begin{equation}
R_{nmkl}=\delta_{nm}\sum_r\Gamma^{(+)}_{nrrk}+\delta_{nk}
\sum_r\Gamma^{(-)}_{lrrm}-\Gamma^{(+)}_{lmnk}-\Gamma^{(-)}_{lmnk},
\end{equation}
in terms of rates given by the golden rule expression 
\begin{eqnarray}
\Gamma^{(+)}_{lmnk}&=&\int_0^{\infty}dte^{-i\omega_{nk}t}{\rm Tr}_B
\left[\tilde{H}_I(t)_{lm}\tilde{H}_I(0)_{nk}\rho_B\right],
\label{Eq:ratesGammaPlus}\\
\Gamma^{(-)}_{lmnk}&=&\int_0^{\infty}dte^{-i\omega_{lm}t}{\rm Tr}_B
\left[\tilde{H}_I(0)_{lm}\tilde{H}_I(t)_{nk}\rho_B\right],
\label{Eq:ratesGammaMinus}
\end{eqnarray}
with $\tilde{H}_I(t)_{lm}=\langle n|e^{itH_B}H_Ie^{-itH_B}|m\rangle$,
and $(\Gamma^{(+)}_{lmnk})^*=\Gamma^{(-)}_{lmnk}$.

We have already pointed out that the dynamics of an open system cannot
be described as a unitary evolution. However, the mapping describing
the evolution is required to be completely positive \cite{Lindblad-76},
implying $\rho\rightarrow\sum_nO_n\rho O^{\dag}_n$, where $\{O_n\}$ is
a set of linear operators on the reduced state space that satisfy
$\sum_nO^{\dag}_nO_n=1$, such to preserve the trace of $\rho$. In the
framework of Lindblad theory \cite{Lindblad-76}, the master equation
can be cast in the form
\begin{equation}
\dot{\rho}(t)_S=-i[H_S,\rho_S(t)]+\frac{1}{2}\sum_j\left\{
[L_j\rho_S(t),L^{\dag}_j]+[L_j,\rho_S(t)L^{\dag}_j]\right\}.
\end{equation}
The Lindblad operators $L_j$ describe the effect of the environment in
the Born-Markov approximation.

\subsubsection{Two level systems and Bloch equations}

The aim of this review is to provide a overview on the mechanisms that
affect qubit dynamics and induce decoherence in solid state
realizations of qubits. Therefore we concentrate on two level systems
and their coupling to the surrounding environment.

The density operator of a two state system is a two dimensional
positive Hermitian operator with trace one. It can thus be expressed
in terms of a basis of Hermitian operators given by the
three Pauli operators
$\boldsymbol{\sigma}=(\sigma_x,\sigma_y,\sigma_z)$ and the $2\times 2$
identity,
\begin{equation}
\rho=\frac{1}{2}(1+{\bf p}\cdot\boldsymbol{\sigma}),
\qquad {\bf p}={\rm Tr}[\rho\boldsymbol{\sigma}]=
\left(\begin{array}{c}
\rho_{01}+\rho_{10}\\i(\rho_{01}-\rho_{10})\\\rho_{00}-\rho_{11}
\end{array}\right).
\end{equation}  
The vector ${\bf p}$ is known as the Bloch vector, and for a spin-1/2
object it represents the expectation values of the spin components
${\bf p}/2\equiv\langle{\bf S}\rangle={\rm Tr}[{\bf S}\rho]$, where
${\bf S}=\boldsymbol{\sigma}/2$, with $\sigma_z$ diagonal in the
$|0\rangle$ $|1\rangle$ basis, $\sigma_z|0\rangle=|0\rangle$ and
$\sigma_z|1\rangle=-|1\rangle$.  
Combining the last equation with the Redfield equation
(\ref{Eq:Redfield-eq}) in the case that $n,m,k,l=0,1$, the master
equation within the Born-Markov approximation for the density matrix 
of a two level system can be expressed as a first order time
differential equation for the expectation value of the spin component
$\langle{\bf S}\rangle= 
(\langle S_x\rangle,\langle S_y\rangle,\langle S_z\rangle)$,
\begin{equation}\label{Eq:Bloch-eq}
\langle\dot{\bf S}\rangle=\boldsymbol{\omega}\times
\langle{\bf S}\rangle-R\langle{\bf S}\rangle+
\langle{\bf S}_0\rangle,
\end{equation}
with $\boldsymbol{\omega}=(0,0,\omega_{01})$. In case of a spin 1/2
particle in a magnetic field defining the $z$ direction, $\omega_{01}$
represents the Zeeman splitting. The inhomogeneous term $\langle{\bf
  S}_0\rangle$ and the relaxation matrix $R$ depend on the rates
Eqs.~(\ref{Eq:ratesGammaPlus}), (\ref{Eq:ratesGammaMinus}). If
$\omega_{01}\gg R_{nmkl}$, it is possible to make a secular
approximation, retaining only terms $R_{nmkl}$ with $n-m=k-l$,
\cite{Redfield-57}, such that the Redfield tensor can be approximated
by the diagonal form
\begin{equation}
R\approx\left(\begin{array}{ccc}
T_2^{-1} & 0 & 0\\
0 & T_2^{-1} & 0\\
0 & 0 & T_1^{-1}\end{array}\right),
\end{equation}
where the relaxation time $T_1$ and the decoherence time $T_2$ 
are given by
\begin{eqnarray}
\frac{1}{T_1}&=&2{\rm Re}(\Gamma_{0110}^{(+)}+\Gamma_{1001}^{(+)}),\\
\frac{1}{T_2}&=&
\frac{1}{2T_1}+\frac{1}{T_{\phi}},\label{Eq:T2}
\end{eqnarray}
with the pure dephasing time $T_{\phi}$, given by
\begin{equation} 
\frac{1}{T_{\phi}}= 
{\rm Re}(\Gamma_{0000}^{(+)}+\Gamma_{1111}^{(+)}
-2\Gamma^{(+)}_{0011}).
\end{equation}
For a system-environment coupling given by a simple bilinear form
$H_I={\cal O}_S\otimes{\cal X}_B$, with ${\cal O}_S$ an operator
acting in the system space ${\cal H}_S$, and ${\cal X}_B$ an operator
acting in the environment space ${\cal H}_B$, the relaxation and
dephasing times $T_1$ and $T_{\phi}$ can be written as    
\begin{eqnarray}
\frac{1}{T_1}&=&4|\langle 0|{\cal O}_S|1\rangle|^2
J(\omega_{01})\coth\frac{\omega_{01}}{2k_BT},\label{Eq:T1}\\
\frac{1}{T_{\phi}}&=&|\langle 0|{\cal O}_S|0\rangle-
\langle 1|{\cal O}_S|1\rangle|^2\left.\frac{J(\omega)}{\omega}
\right|_{\omega\rightarrow 0}2k_BT,\label{Eq:Tphi}
\end{eqnarray}  
where the spectral density $J(\omega)$ is the Fourier transform of the
environment time correlator
\begin{equation}
J(\omega)=\int_{-\infty}^{\infty}dt{\rm Tr}_B\left[{\cal X}_B{\cal
  X}_B(t)\rho_B\right] e^{-i\omega t}.
\end{equation}

The first term in Eq.~(\ref{Eq:Bloch-eq}) produces a rotation of the
Bloch vector along the $z$ direction. If $R=0$ we have
the classical picture of a magnetic moment precessing along the
externally applied magnetic field. The second term proportional to $R$
describes an exponential damping of the component of the Bloch
vector. $T_1$ describes the decay of the longitudinal component
of the Bloch vector, while $T_2$ describes the decay of the transverse
component. 

We remark that the Markovian results Eq.~(\ref{Eq:T2}) satisfy the
expected fundamental Korringa relation \cite{Abragam-61}.

\subsection{Spin-boson model}

\label{Sec:Spin-boson}

Here we describe a simple model to treat the dynamics of a
two-level system in contact with a reservoir. We consider a generic
two-level system described by the Hamiltonian
\begin{equation}\label{Eq:qubit-Ham-SpinBoson}
{\cal H}_S=\frac{\Delta}{2}\sigma_x+\frac{\epsilon}{2}\sigma_z.
\end{equation}
In order to include the effect of dissipation in the quantum
formalism, it is customary to follow the Caldeira-Leggett
\cite{CaldeiraLeggett83,Zwerger-RMP87,Weiss99} approach.  
A bath of harmonic oscillators at thermal equilibrium at
temperature $T$ is introduced to describe the degrees of freedom
of the environment. The system+bath Hamiltonian is
\begin{eqnarray}
H&=&H_S+H_B+H_{SB},\\
H_B&=&\frac{1}{2}\sum_{\alpha}\omega_{\alpha}\left(
b^{\dag}_{\alpha}b_{\alpha}+\frac{1}{2}\right),\\
H_{SB}&=&{\cal O}_S\otimes{\cal X}_B=
\sigma_z\sum_{\alpha}c_{\alpha}
\left(b_{\alpha}+b^{\dag}_{\alpha}\right), 
\end{eqnarray}   
where ${\cal H}_S$ is the quantized Hamiltonian of the system
Eq.~(\ref{Eq:qubit-Ham-SpinBoson}), ${\cal
  H}_B$ is the bath Hamiltonian, described by independent bosonic
degrees of freedom with frequencies $\omega_{\alpha}$. The
coupling between the system and the bath degrees of freedom is
described by ${\cal H}_{SB}$, where ${\cal O}_S=\sigma_z$, 
${\cal X}_B=\sum_{\alpha}c_{\alpha}(b_{\alpha}+b^{\dag}_{\alpha})$,
and  $c_{\alpha}$  are coupling parameters. 

A rigorous treatment of the spin-boson model in the Born approximation
without making use of the Markov approximation is presented in
\cite{LossDiVincenzo-PRB05,LossDiVincenzo-condmat03}.  
The eigenstates of the Hamiltonian
Eq.~(\ref{Eq:qubit-Ham-SpinBoson}) are 
\begin{eqnarray}
|0\rangle&=&\frac{1}{\sqrt{2}}\left(\sqrt{1+\frac{\epsilon}{\omega_{01}}} 
|+\rangle+\sqrt{1-\frac{\epsilon}{\omega_{01}}}|-\rangle\right),\\
|1\rangle&=&\frac{1}{\sqrt{2}}\left(\sqrt{1-\frac{\epsilon}{\omega_{01}}} 
|+\rangle-\sqrt{1+\frac{\epsilon}{\omega_{01}}}|-\rangle\right), 
\end{eqnarray}
where $|\pm\rangle$ are eigenstates of $\sigma_z$,
$\sigma_z|\pm\rangle=\pm|\pm\rangle$, and
$\omega_{01}=\sqrt{\Delta^2+\epsilon^2}$. The initial state of a
system of quantum harmonic oscillators in thermal equilibrium is 
\begin{equation}
\rho_B={\cal Z}_B^{-1}\exp(-\beta{\cal H}_B), 
\qquad {\cal Z}_B={\rm Tr}\exp(-\beta{\cal H}_B).
\end{equation}
All the informations of the bath, such as the bath frequencies
$\omega_{\alpha}$ and the coupling parameters $c_{\alpha}$ appearing
in the Hamiltonian, are contained in the spectral density $J(\omega)$
of the system-bath coupling, 
\begin{equation}
J(\omega)
=\frac{\pi}{2}\sum_{\alpha}c_{\alpha}^2
\delta(\omega-\omega_{\alpha}).  
\end{equation}
Here we limit our attention to the Markovian case, and make use of the
general Redfield theory described in the previous section. From the
formula Eqs.~(\ref{Eq:T1}), (\ref{Eq:Tphi}), the
relaxation and dephasing rates take the form 
\begin{eqnarray}
\frac{1}{T_1}&=&\left(\frac{\Delta}{\omega_{01}}\right)^2 
J(\omega_{01})\coth\frac{\omega_{01}}{2k_BT},\\
\frac{1}{T_{\phi}}&=&\left(\frac{\epsilon}{\omega_{01}}\right)^2
\left.\frac{J(\omega)}{\omega}
\right|_{\omega\rightarrow 0}2k_BT.
\end{eqnarray}

\subsection{Spin qubits}

Per antonomasia, the two-state system that nature provides us with is the
intrinsic angular momentum of the electron: the spin 1/2. It is therefore
natural to choose the electron spin as the two-state system that
encodes the qubit. The spin of the electron can have much longer
decoherence time than the charge degrees of freedom. Nevertheless,
isolating the spin degree of freedom of an electron to a degree
required for quantum computation is not at all an easy task. Moreover,
in order to be used for quantum 
computational purposes, electron spin-based qubits must be designed as
scalable devices that can be externally controlled, coupled,
manipulated, and read-out, i.e. they must satisfy the DiVincenzo
criteria \cite{DiVincenzo-criteria}.   
A successful and promising device for the physical implementation
of electron spin-based qubits is the semiconductor quantum dot
\cite{LossDiVincenzo-PRA98}.   

\subsection{Semiconductor quantum dots}

The quantum dots owe their name to the zero-dimensional character of
such devices. Can be considered as a quantum box that can be filled
with electrons (or holes) which occupy the available discretized
states of the system. The electrons can tunnel on and off the dot, 
which is coupled to large reservoir via tunnel barriers. The
height of the barriers, and consequently the rates for tunneling
through the barriers on and off the dot, can be controlled via the 
application of gate electrodes. Electrostatic gates can also be used
to tune the electrostatic potential of the dot with respect to the
reservoirs, such that the ladder of energy levels in the dot can be
shifted up or down with respect to the energy of the reservoir. 
External bias voltages can be applied and transport properties can be
measured.  
  
Quantum dots are basically characterized by the quantized level
structure, for which they are considered as artificial atoms,
and by the transport state of the dot, that can be active or blocked,
and depends on the combination of bias and gate voltages applied. In
fact, the Coulomb repulsion between the electrons in the dot
determines an energy cost for adding an extra electron in the dot. At
low temperatures, the tunneling of electrons on and off the dot can
be drastically suppressed, and the dot is in the so called Coulomb
blockade.   

\begin{figure}
 \begin{center}
 \includegraphics[width=6cm]{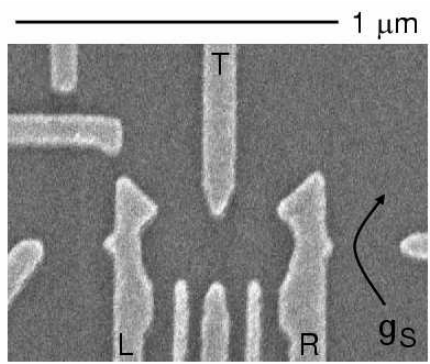}
    \caption{Scanning electron micrograph of a double quantum dot. 
      \cite{Petta-arXiv0709.0920}
      (With permission from the Authors).
    \label{Fig:DoubleQD-Petta}}
 \end{center}
\end{figure}

Many kinds of quantum dots have been realized so far. Here we focus
the attention on lateral III-V semiconductor quantum dots, as those in
Fig.~\ref{Fig:DoubleQD-Petta}. These
devices are fabricated from heterostructures of GaAs and AlGaAs grown
by molecular  beam epitaxy. The energy potential along the growth
direction of such a structure has a minimum at the
interface of the two layers, which is also asymmetric with respect to
the growth direction. Free electrons are introduced by doping the
AlGaAs layer with Si, which accumulate at the GaAs/AlGaAs interface,
deep down in the minimum of the vertical potential, that provides
strong confinement of the electrons along the growth direction. 
At the same time, the electrons are free to move along the interface,
where they form a two dimensional electron gas (2DEG), that can have a
high mobility and a relatively low electron density (typically
$10^5-10^7~{\rm cm}^2/{\rm Vs}$ and $\approx 10^{15}~{\rm m}^{-2}$). The
low density results in a relatively long Fermi wavelength ($\approx
40~{\rm nm}$) and a large screening length, such that via application
of an electric field, obtained through metal gate electrodes on top of
the heterostructure negatively charged, the 2DEG can be locally
depleted. Therefore, by suitable designing the gate structure it is
possible to isolate small islands of the 2DEG, thus creating a
dot. When the lateral size of the dot is compared to the Fermi
wavelength, the energy level structure of the dot becomes discretized,
and at temperatures down to tens of mK, the energy separation of the
levels becomes much higher than the temperature, such that quantum
phenomena start to play a significant role.    

\subsection{Spin relaxation and spin dephasing mechanisms in
  quantum dots}

The electron spin in semiconductor quantum dots can be isolated and
controlled with a high accuracy, but it still suffers from decoherence
due to the unavoidable coupling with the surrounding environment. In
order to implement quantum computation algorithms with electron
spin-based qubits in semiconductor quantum dots, it is necessary to
engineer the devices in such a way to preserve the coherence of the
electron spin states for sufficiently long time scales. Besides the
fundamental interest, it is therefore important to theoretically
understand which sources of dissipation and decoherence affect the
electron spin in quantum dots, and to find ways to reduce  their
influence on the spin-qubit dynamics as much as possible.   

Two kinds of environment turn out to mainly affect the dynamics of an
electron spin in a quantum dot, the phonons in the lattice, and the
spins of atomic nuclei in the quantum dot.

Phonon-induced relaxation in semiconductor quantum dots has attracted
some attention from a theoretical point of view for the first time in
\cite{Khaetskii-PRB00,Khaetskii-PRB01}.   
The lattice phonons do not couple directly to the spin degree of
freedom. However, even without the application of external electric
fields, the breaking of inversion symmetry in GaAs
gives rise to spin-orbit interaction,
that couples the spin  and the orbital degrees of freedom. The
latter, being coupled to the phonons, provide an indirect coupling
between the electron spin and the phonons, that constitute a large
dissipative bosonic reservoir and provide a source of decoherence and
relaxation.  Short time correlations in the phonon bath induce a
Markovian dynamics of the electron spin, with well defined relaxation
and decoherence time $T_1$ and $T_2$. It turns out that
effectively the phonon-induced pure dephasing time $T_{\phi}$ of an
electron spin in a quantum dot in the presence of a magnetic field
diverges. In the Bloch picture, pure dephasing arise from longitudinal
fluctuations of the magnetic field, while a perturbative treatment of
the spin-orbit interaction gives rise, within first order, to a
fluctuating magnetic field perpendicular to the applied one. As a
consequence the decoherence time $T_2$ is limited only by its upper
bound $T_1$, $T_2=2T_1$. In turn the relaxation time $T_1$ shows a
strong dependence on the magnetic field, $T_1\propto B^5$, that has
been confirmed experimentally, where a very long relaxation time up to
$T_1\approx 1~{\rm s}$ has been measured for a magnetic field of
$B=1{\rm T}$ \cite{Amasha-arXiv07}.     

Hyperfine interaction was for the first time taken into consideration as
a source of decoherence for an electron spin confined in a quantum dot
in \cite{Burkard-PRB99}. In GaAs there are $\approx 5\times 10^{21}$
atoms in $1~{\rm cm}^3$. Therefore, the linear extension of a typical
GaAs quantum dot, that is of order of the Fermi wavelength $\approx
40~{\rm nm}$, encompasses roughly 200 atoms, from which it can be
estimated that the wavefunction of an electron in a GaAs quantum dot
overlaps with $\approx 10^5$ nuclei. The electron spin and the nuclear
spins in the dot couple  via the Fermi contact hyperfine interaction, 
that creates entanglement between them and strongly affect the electron
spin dynamics. It turns out that long time correlations in the nuclear
spin system induce a non-Markovian dynamics of the electron spin,
with non exponential decay in time of the expectation values of the
electron spin components. In a large applied magnetic field $B$, the
dynamics in the nuclear field due to the hyperfine interaction can be
treated perturbatively and it turns out that flip-flop dynamics starts
to affect the nuclear field in a time that scales like the number of
nuclear spins, $\propto N$. For shorter times the nuclear field is
static and the transverse component shows a Gaussian decay, that is
due to the statistical distribution of nuclear spin
states.

We remark that the phonon induced relaxation rate of the electron spin
is enhanced by an applied magnetic field, whereas the influence of the
hyperfine interaction is reduced by a large Zeeman splitting.

\subsection{Hyperfine-induced decoherence in spin qubits}

\subsection{Hyperfine interaction}

The spin of an electron and the atomic nuclear spin can interact
through the hyperfine Fermi contact interaction, a spin-spin
interaction that takes place when the electron and the nucleus occupy
the same position in space, from which the term ``contact''. The
origin of the hyperfine coupling can be understood considering the
electromagnetic 
interaction of an electron with the magnetic field produced by a
nucleus. Without loss of generality the magnetic properties of a
nucleus can be described as those of a magnetic dipole
$\boldsymbol{\mu}_N=\mu_N\hbar{\bf I}$, where $\mu_N$ is the nuclear
magneton, and ${\bf I}$ is the nuclear spin operator. The interaction
of a nuclear dipole $\boldsymbol{\mu}_N$ with the electronic shell
gives a rather small effect, and can be treated using a perturbative
method. In the non-relativistic Pauli description of the electron, the 
Hamiltonian of an electron in a magnetic field ${\bf B}={\bf
  \nabla}\times{\bf A}$ produced by a vector
potential ${\bf A}$ is given by
\begin{equation}\label{Eq:Ham-e-inBfield}
{\cal H}=\frac{1}{2m}\left({\bf p}+\frac{e}{c}{\bf A}\right)^2+
2\mu_B{\bf S}\cdot({\bf \nabla}\times{\bf A}),
\end{equation} 
where ${\bf S}$ is the electron spin operator.
The vector potential produced by a magnetic dipole $\boldsymbol{\mu}$
at position ${\bf r}$ is, according to classical electromagnetism,
${\bf A}=(\boldsymbol{\mu}\times{\bf r})/r^3={\bf \nabla}\times
(\boldsymbol{\mu}/r)$. Neglecting the term quadratic in the vector
potential and replacing $\hbar{\bf L}={\bf r}\times{\bf p}$ for the
electron orbital momentum operator,
the Hamiltonian Eq.~(\ref{Eq:Ham-e-inBfield}) can be written as
\begin{equation}\label{Eq:Ham-spin-mu}
{\cal H}=
2\mu_B\frac{{\bf L}\cdot\boldsymbol{\mu}}{r^3}+
2\mu_B({\bf S}\cdot{\bf \nabla})
(\boldsymbol{\mu}\cdot{\bf \nabla})\frac{1}{r}-2\mu_B
({\bf S}\cdot\boldsymbol{\mu})\nabla^2\frac{1}{r}.
\end{equation}
The magnetic interaction of the nuclear spin and the electron spin is
contained in the second and the third term of
Eq.~(\ref{Eq:Ham-spin-mu}), and it is obtained after integration over
the orbital degrees of freedom, i.e. it has to be understood as
applied to an electron orbital state $\psi_{\rm el}({\bf r})$. 
For ${\bf r}\ne 0$, the terms involving the electron spin
${\bf S}$ in Eq.~(\ref{Eq:Ham-spin-mu}) behave regularly, the last
term vanishes identically, while the second term produces a usual
dipole-dipole interaction $2\mu_B[3({\bf S}\cdot{\bf r})
(\boldsymbol{\mu}\cdot{\bf r})/r^5-{\bf
  S}\cdot\boldsymbol{\mu}/r^3]$. 
The case ${\bf r}=0$ needs to be treated more carefully. It can be
shown \cite{Abragam-61} that the dominant contribution of the spin
dependent part of Eq.~(\ref{Eq:Ham-spin-mu}) reduces to
$(16\pi/3)\mu_B({\bf S}\cdot\boldsymbol{\mu})\delta({\bf r})$, and
once applied on the electron orbital wave function is given by
\begin{equation}
{\cal H}_{\rm hy}=\frac{16}{3}\pi\mu_B
|\psi_{\rm el}(0)|^2{\bf S}\cdot\boldsymbol{\mu},
\end{equation}
which is finite for $s$ electrons and zero for others. The Hamiltonian
for the magnetic interaction of the electron with the nucleus can be
written as
\begin{equation}
{\cal H}=
2\mu_B\mu_N\hbar{\bf I}\cdot\left[\frac{\bf L}{r^3}-\frac{\bf S}{r^3}
+3\frac{{\bf r}({\bf S}\cdot{\bf r})}{r^5}+\frac{8}{3}
\pi{\bf S}\delta({\bf r})\right].
\end{equation} 

\subsubsection{Hyperfine interaction in semiconductor quantum dots}

\begin{figure}
 \begin{center}
 \includegraphics[width=8cm]{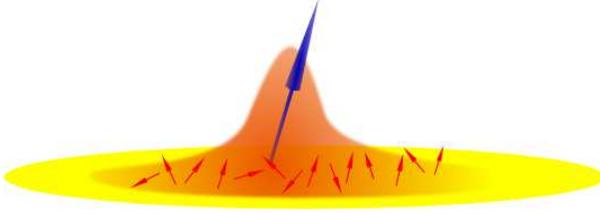}
    \caption{Schematic representation of the orbital wave function of
      an electron in a quantum dot. Due to the spatial extension of
      the wave function, the electron spin (big blue arrow) couples to
      many nuclear spins (small red arrows).
    \label{Fig:Hyperfine-int-QD}}
 \end{center}
\end{figure}
In a semiconductor quantum dot an electron is confined in a 2D region
of space whose linear extension is of the order of the Fermi wavelength, 
for GaAs about $\sim 100 {\rm nm}$, that is much larger than the
typical lattice spacing of the crystal ($\sim$\AA). As a result
a discretization of the energy levels in the dot appears, with an orbital 
level spacing that, for lateral quantum dot containing single electrons,
is  much greater than the typical energy scale of the hyperfine
interaction. As opposed to the case of single atoms, the electron
orbital wave function in a quantum dot extends over a region much
larger than the lattice size, such that the electron spin couples to
many nuclear spins, as schematically shown in
Fig.~\ref{Fig:Hyperfine-int-QD}. 
The effective hyperfine Hamiltonian describing the interaction of a
single electron with the nuclei in the dot can be written as 
\begin{equation}\label{Eq:HyHam}
{\cal H}={\bf S}\cdot{\bf h},\qquad 
{\bf h}=Av_0\sum_{k=1}^N|\psi_0({\bf r}_k)|^2{\bf I}_k
\equiv\sum_{k=1}^{N}A_k{\bf I}_k,
\end{equation} 
where ${\bf S}$ is the electron spin operator, ${\bf h}$ is the
so-called Overhauser field, given by the sum of all the ${\bf I}_k$
nuclear spin operators, weighted by the position dependent coupling
strength $A_k=v_0A|\psi_0({\bf r}_k)|$, where the square modulus of
the electron envelope wave function at the $k$th lattice
site. Typically the electron can be assumed to be in the quantum 
dot orbital ground state. $v_0$ is the volume of the crystal unit cell
containing one nuclear spin and $A=16\pi\mu_B\mu_N\hbar/3$ is the
contact hyperfine coupling strength. In GaAs the nuclear spin is
$I=3/2$ and an estimate of the interaction strength, weighted by the
abundances of the three isotopes naturally present (${}^{69}{\rm
  Ga}$, ${}^{71}{\rm Ga}$, and ${}^{75}{\rm Ga}$), yields $A\approx
90\mu{\rm eV}$.  

The inhomogeneity of the electron wave function results in a nonuniform
hyperfine coupling strength $A_k$, that depends on the probability to
find the electron in the nuclear lattice site $k$, resulting in a
subtle and complex many-body quantum mechanical behavior, with the
nuclear spin affecting the electron spin time evolution, and the
electron spin acting back on the dynamics of each of the nuclei in
turn. 

From the point of view of the electron spin, entanglement with
the degrees of freedom of nuclear spins arising from the
hyperfine coupling constitutes a decoherence mechanism.

Since the Overhauser field ${\bf h}$ appearing in Eq.~(\ref{Eq:HyHam})
is composed by the sum of a large number of spins, it is natural to
question whether the nuclear field can be approximated to a
classical object and to which extent this approximation gives correct
results. In a relatively recent work by Coish {\it et al.} 
\cite{Coish-JAppPhys06} it has been theoretically shown that, for the
special case of a uniform hyperfine coupling constants $A_k=A/N$,
arising from a constant wave function in the dot
$\psi_0=1/\sqrt{v_0N}$, the dynamics obtained in the mean field
approximation and the quantum evolution show agreement up to the
transverse-spin correlation time 
$\tau_c$, which diverges in the zero external magnetic field case
(unphysical result due to the assumption of constant coupling), but
that saturates to a finite value in case of a finite external magnetic
field.

\subsubsection{Fluctuation timescales of the nuclear field}

The nuclear field is quantum many-body interacting spin system whose
field orientation and magnitude change over time. This change is due
to the combined effect of the inter-nuclear dipole-dipole interaction
and the hyperfine interaction between electron and the nuclei. The
dipolar interaction does not conserve the total nuclear spin and thus
can be responsible for changes in the nuclear spin configuration. Those
changes, combined with the spatial variation of the hyperfine coupling
constant, lead to a different value of the nuclear field seen by the
electron spin and thus to its decoherence. Here we briefly outline the
timescales in which those mechanisms take place, in order of
decreasing timescales.

The strength of the effective magnetic dipole-dipole interaction
between neighboring nuclei in GaAs is directly given by the width of
the nuclear magnetic resonance (NMR) line to be $\sim(100\mu s)^{-1}$
\cite{Paget-PRB77} and its inverse can be taken as an estimate for the
timescale in which a change in the nuclear configuration due to
dipolar interaction takes place, i.e. $T_{d-d}\approx 100\mu s$, which
is just the period of precession of a nuclear spin in the local
magnetic field generated by its neighbors. This timescale is so long
that a great number of other decoherence mechanisms start to
play a significant role before nuclear dipole effects start to matter. 

Besides spin diffusion driven by nuclear dipole-dipole interaction,
the nuclear field can undergo a change due to the flip-flop term in
the Hamiltonian Eq.~(\ref{Eq:HyHam}). In a large external field $B$,
the flip-flop term can be treated within the framework of perturbation
theory, as it will be explained in the next section. We anticipate
here that the hyperfine mediated dynamics in the nuclear field has a
timescale given by $\propto A/N$. This means that up to this
timescale the nuclear field can be considered as static.

\subsection{Decoherence due to hyperfine-induced electron 
spin dynamics}  

An early treatment of the hyperfine interaction as a decoherence
mechanism for single electron spins confined in quantum dots whose
carried out in \cite{Burkard-PRB99}. There, a second order
time-dependent perturbation expansion of the hyperfine interaction in
a magnetic field was performed with respect to the flip-flop
transverse term $A(h_+S_-+h_-S_+)/2$ for a constant hyperfine coupling
$A$ and a long-time longitudinal spin-flip probability $\sim 1/p^2N$
is obtained, where $p$ is the nuclear spin polarization. As a result
beside a large external magnetic field, a large polarization $p$ and a
large number of nuclei in the dot would suppress the spin-flip
probability. 

The first signature of the non-Markovian behavior of the nuclear spin
bath appeared in \cite{Khaetskii-PRL02}, where an exact solution for
the fully polarized case $p=1$ was provided. In that case the
decoherence is due to a non-uniform hyperfine coupling that depends on
the probability for the electron to be located at different nuclear
sites. A remarkable feature of the non-Markovian behavior is the
long-time power law decay of the electron spin correlator, $\sim
1/t^{3/2}$, in strong Zeeman field, according to which the
longitudinal electron spin component decays of a fraction of $\sim
1/N$, in a time $\sim N/A$.

\subsubsection{Single-electron spin decoherence in large Zeeman splitting}

A detailed and comprehensive treatment of the hyperfine interaction
\cite{Coish-PRB04} provides an analytical result for the electron spin
dynamics for arbitrary nuclear spin $I$ and nuclear polarization
$p$. A generalized master equation (GME) approach allows a treatment of
the transverse electron spin-nuclear spin flip-flop terms in the
Hamiltonian with an external field in a well controlled perturbative
way. An expansion of the self-energy in the exact Nakajima-Zwanzig GME
shows a rich electron spin dynamics, with exponential and
non exponential decaying contributions and undamped oscillations. The
form of the decay of the transverse and longitudinal electron spin
component is obtained in high magnetic field up to forth order in 
perturbation theory. 

The hyperfine Hamiltonian in an external magnetic field is
\begin{equation}\label{Eq:HpHam-B}
{\cal H}=bS_z+\epsilon_{nz}I_z+{\bf h}\cdot{\bf S},
\end{equation} 
where $b=g^*\mu_BB_z$ ($\epsilon_{nz}=g_I\mu_NB_z$) is the electron
(nuclear) Zeeman splitting in a magnetic field defining the $z$-axis
$B_z$, $g^*$ ($g_I$) the effective electron (nuclear) $g$-factor, and
$\mu_B$ ($\mu_N$) the Bohr (nuclear) magneton. In the rotating frame
with respect to the nuclear Zeeman term the Hamiltonian can be
separated into a longitudinal (unperturbed) and transverse
(perturbation) term,
\begin{equation}\label{Eq:HamiltLargeZeeman}
{\cal H}=\underbrace{(b+h_z)S_z}_{{\cal H}_0}+
\underbrace{(h_+S_-+h_-S_+)/2}_{V}
\end{equation}
In absence of $V$, $\langle S_z\rangle_t$ is constant, since
$[{\cal H}_0,S_z]=0$, but the transverse component $\langle
S_{\pm}\rangle_t$ evolves in time in a non trivial way. For a large
number of nuclear spins $N\sim 10^5$ (GaAs dot) a direct
application of the central limit theorem gives a Gaussian distribution  
for the eigenvalues of $h_z$ with mean $h_0=\langle
h_z\rangle$ and variance
$\sigma\approx A/\sqrt{N}$. The transverse correlator for an
initial state given by the product state of the initial electron spin
state $\rho_S(0)$ and incoherent Gaussian distributed nuclear mixture
state is 
\begin{equation}\label{Eq:Gauss-decay}
\langle S_+\rangle_t\approx\langle S_+
\rangle_0\exp[-t^2/2\tau^2+i(b+h_0)t],\qquad 
\tau=\frac{1}{\sigma}=\frac{2\hbar}{A}\sqrt{\frac{N}{1-p^2}}.
\end{equation}
Choosing as nuclear initial state the pure state
$|\psi_I(0)\rangle=\prod_j(\sqrt{1+p}|\uparrow_j\rangle
+e^{i\phi_j}\sqrt{1-p}|\downarrow_j\rangle)/\sqrt{2}$, for a certain
polarization $p$, the same result Eq.~(\ref{Eq:Gauss-decay}) with
$h_0=pN$ comes out. 

The reason for this decay lies in the choice of the initial nuclear
state containing many $h_z$ eigenstates and can also be obtained
choosing the nuclear field in a $h_z$ eigenstate, but with the
electron spin in a transverse initial state. This decay is 
reversible and can be removed with a standard spin echo
technique \cite{Petta-Science05, Koppens-PRL07}. Such an experiment
therefore reveals only the decay 
due to the transverse flip-flop term $V$ Eq.~(\ref{Eq:HpHam-B}). A
procedure more suitable for a quantum computation algorithm would be a
strong Von Neumann measurement of the nuclear field that would then
prepare a $h_z$ eigenstate, leading to simple precession with no decay
\cite{Klauser-PRB06,Stepanenko-PRL06}.   

Analysis of the GME in the Born approximation for a very high magnetic
field ($b\gg N$) provides an asymptotic form to leading orders in $\sim
1/\omega_n=1/(b+h^z_n)$, 
\begin{equation}
\langle S_+\rangle_t\approx\sigma_+^{\rm osc}(t)+\sigma_+^{\rm
  dec}(t),\qquad
\langle S_z\rangle_t\approx\langle S_z\rangle_{\infty}+\sigma_z^{\rm
  dec}(t),
\end{equation}
where $\langle S_z\rangle_{\infty}\propto\langle S_z\rangle_0$, 
$\sigma_+^{\rm osc}(t)\propto\langle S_+\rangle_0e^{i\omega_nt}$, and 
$\sigma_{+/z}^{\rm dec}(t)\propto\delta/t^{3/2}$ for a parabolic
confinement in the dot, with $\delta\ll 1$. Even for a $h_z$
eigenstate, for which no decay is expected in zeroth order in the
transverse electron nuclear-spin flip-flop interaction, a long time
irreversible decay takes place, that is due to the spatial
variation of the hyperfine coupling constant.

\subsubsection{Single-spin ESR: universal phase shift and power law decay}

\begin{figure}
 \begin{center}
 \includegraphics[width=7cm]{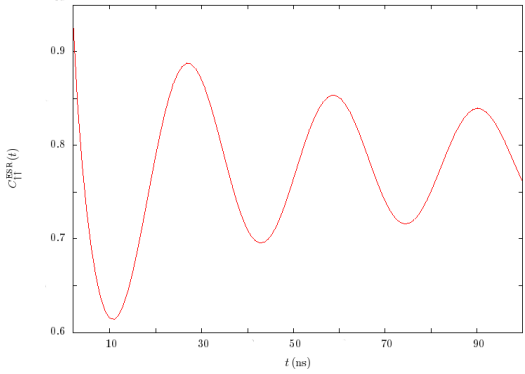}
    \caption{Decay of the driven Rabi oscillation in ESR showing a
      power law decay $\propto 1/\sqrt{t}$ and a universal phase shift of
      $\pi/4$, as given by Eq.~(\ref{Eq:ESRpowerlaw}), valid for
      $t\gg{\rm max}(1/\sigma,1/b_{\rm ac},b_{\rm ac}/2\sigma^2)$. For
      the plot the value $b=\sigma=0.4~{\rm Ghz}$ have been chosen.
    \label{Fig:powerlaw}}
 \end{center}
\end{figure}
Here we describe the situation in which the electron spin is
coherently driven via pulsed magnetic resonance, while coupled to a
nuclear long-time correlated spin bath. Recent remarkable experimental  
results \cite{Koppens-PRL07} show a coherent electron spin 
oscillation, even for a Rabi period much longer than $T_2^*=10-20~{\rm
  ns}$. A non-exponential decay of the Rabi oscillations is observed,
obeying a power law decay with the appearance of a universal phase
shift. 

Consider a quantum dot in a time independent magnetic field defining
the $z$ direction. In addition, an oscillating magnetic field is
applied in the plane, along the $x$ direction. 
For a large number of nuclei in the dot ($N\sim 10^6$ in GaAs dots)
the field $h_z$ is Gaussian distributed, with mean $h_0$ and variance
$\sigma$ \cite{Merkulov-PRB02,Khaetskii-PRL02,Coish-PRB04}. In the
case of strong external field ($b\gg\sigma$, with $b=g^*\mu_BB_z$),
neglecting transverse electron-nuclear spin flip-flop terms in the
hyperfine interaction, the Hamiltonian is ($\hbar=1$) 
\begin{equation}\label{Eq:ESRham}   
{\cal H}=(b+h_z)S_z+b_{\rm ac}\cos(\omega t)S_x,
\end{equation}
where $b_{\rm ac}=g^*\mu_BB_{\rm ac}$, $\omega$ and $B_{\rm ac}$ being
the frequency and amplitude of the ESR driving field. $h_z$ is
considered static (justified for $t<1\mu{\rm s}$), and the
assumption $\omega=b+h_0$ is made.  
In the rotating wave approximation (valid for $(b_{\rm ac}/b)^2\ll
1$), after averaging over the Gaussian distribution of $h_z$, the decay
of the driven Rabi oscillation is given by \cite{Koppens-PRL07} 
\begin{equation}\label{Eq:ESRpowerlaw}
P_{\uparrow}^{\rm ESR}(t)\sim 1-C+\sqrt{\frac{b_{\rm ac}}{8\sigma^2t}}
\cos\left(\frac{b_{\rm ac}}{2}t+\frac{\pi}{4}\right)
+{\cal O}\left(\frac{1}{t^{3/2}}\right),
\end{equation}
for $t\gg{\rm max}(1/\sigma,1/b_{\rm ac},b_{\rm ac}/2\sigma^2)$, with
$C=\exp(b_{\rm ac}^2/8\sigma^2){\rm erfc}(b_{\rm
  ac}/\sqrt{8}\sigma)\sqrt{2\pi}b_{\rm ac}/8\sigma$.
The remarkable features appearing in the experiment are the $\sim
1/\sqrt{t}$ power law decay and the universal $\pi/4$ phase
shift. The reason for the appearance of these features is that the
nuclear field $h_z$ does not change over a timescale much longer
than the Rabi period. Since different values of $h_z$ determine 
different oscillation frequencies, an average over the distribution in
$h_z$ give rise to a decay in the coherence of the driven electron
spin, and the off-resonant contributions also determine the phase
shift. The fact that coherent Rabi oscillations are visible even when
the Rabi period is much longer than the transverse spin decay time
$\tau\sim 15 {\rm ns}$ has its origin in the fact that the power law
decay sets in already after a short time $1/\sigma\sim 15 {\rm ns}$.  

In order to measure the electron spin state in the experiment 
\cite{Koppens-PRL07}, a spin-charge conversion technique 
is implemented by operating a double quantum dot in the spin 
blockade regime \cite{OnoTarucha-PRL04,Koppens-Nature06}, in which the
transport through the dots can occur 
only via transitions from spin states with one electron per dot,
$|(1,1)\rangle$, to the singlet state in the right dot,
$|(0,2)\rangle$. The Pauli exclusion principle, that does not allows
two electron with same 
spin state to occupy the same orbital, allows transport only for
antiparallel spins. Transport of states spin-triplet states is therefore
blocked. The oscillating transverse magnetic field rotates the spins,
therefore unblocking an initial state with even parity spin state
\cite{Koppens-Nature06}.

\subsubsection{Single-triplet decoherence in a double quantum dot}

An alternative way to implement a qubit with electron spin in quantum
dots is to consider a double quantum dot with two spins, one per dot,
and encode the qubit in the subspace with zero
$z$-projection of the total spin $S_{\rm tot}^z=S_1^z+S_2^z=0$. 
Advantages of this scheme is the possibility of reducing the hyperfine
coupling in case of symmetric dots. At the same time additional
decoherence due to the coupling to the orbital degree of freedom and
leakage errors may appear.

The effective Hamiltonian for the one-electron-per-dot configuration
can be written as
\begin{equation}
{\cal H}_{dd}=\epsilon_zS_z+{\bf h}\cdot{\bf S}+
\delta{\bf h}\cdot\delta{\bf S}+\frac{J}{2}{\bf S}\cdot{\bf S}-J,
\end{equation}
where ${\bf S}={\bf S}_1+{\bf S}_2$, $\delta{\bf S}={\bf S}_1-{\bf
  S}_2$,  ${\bf h}={\bf h}_1+{\bf h}_2$, and $\delta{\bf h}={\bf
  h}_1-{\bf h}_2$. Here $J$ is the Heisenberg exchange coupling
between the two electron spins. For definiteness we work in a regime
of large Zeeman splitting due to an external magnetic field,
$\epsilon_z=g^*\mu_B\gg{\rm max}\{\langle\delta{\bf  h}\rangle_{\rm
  rms}, \langle{\bf h}\rangle_{\rm rms}\}$, where $\langle{\cal
  O}\rangle_{\rm rms}=\langle\psi_I|{\cal O}|\psi_I\rangle^{1/2}$
denotes the root-mean-square expectation value of the operator ${\cal
  O}$ on the nuclear state $|\psi_I\rangle$. Requiring $\epsilon_z\gg
J$, where $J$ is taken to be positive without loss of generality, the
large Zeeman splitting condition renders the relevant spin Hamiltonian
block diagonal with respect to the eigensubspaces of $S_z$. In the
$S_z=0$ subspace the spin Hamiltonian for the singlet $|S\rangle$ and
$S_z=0$ triplet $|T_0\rangle$, to zeroth order in the inverse
Zeeman splitting $1/\epsilon_z$, is given by ${\cal H}_0=(J/2){\bf
  S}\cdot{\bf S}+\delta h^z\delta S^z$. The effective qubit
Hamiltonian  in terms of the vector consisting of Pauli matrices
$\boldsymbol{\tau}=(\tau^x,\tau^y,\tau^z)$, with the computational
states $|S\rangle\rightarrow|\tau^z=-1\rangle$ and
$|T_0\rangle\rightarrow|\tau^z=1\rangle$ , has the form
\begin{equation}\label{Eq:Qubit-Ham}  
{\cal H}_0=\frac{J}{2}(1+\tau^z)+\delta h^z\tau^x.
\end{equation}
A systematic treatment of the dynamics induced by the Hamiltonian
Eq.~(\ref{Eq:Qubit-Ham}) can be found in
\cite{Klauser-PRB06,Coish-PRB05}. The eigenstates of ${\cal H}_0$ are
given by a product state between a nuclear eigenstate $|n\rangle$ of
$\delta h^z$ and a superposition of $|S\rangle$ and $|T_0\rangle$,
therefore ${\cal H}_0$ does not lead to any dynamics in the nuclear
field. The correlator $C_{T_0S}$ is defined as 
the probability to find the electron spins in the state $|T_0\rangle$
at time $t>0$, provided the initial state ($t=0$) was 
$|\psi(0)\rangle=|S\rangle\otimes|\psi_I\rangle$, with 
$|\psi_I\rangle$ a superposition of $\delta h^z$ eigenstates,  
\begin{equation}
C_{T_0S}(t)=\sum_n\rho_I(n)|\langle n|\otimes\langle
T_0|e^{-i{\cal H}_0t}|S\rangle\otimes|n\rangle|^2,  
\end{equation} 
where $\rho_I(n)$ diagonal matrix element of
$|\psi_I\rangle\langle\psi_I|$ in the $\{|n\rangle\}$ basis. For a
Gaussian distributed field $\delta h_z$, with mean $x_0$ and variance
$\sigma_0$, the asymptotics of $C_{T_0S}$ saturates to finite value
that deviates from the semiclassical results [$C_{T_0S}^{\rm
  semicl}(\infty)=1/2$] for $J\ll x_0$ \cite{Coish-PRB05}
\begin{equation}
C_{T_0S}(\infty)\sim\left\{\begin{array}{cc}
\frac{1}{2}-\frac{1}{8}\left(\frac{J}{x_0}\right)^2, &
\sigma_0,J\ll x_0,\\
2\left(\frac{x_0}{J}\right)^2, &  \sigma_0\ll x_0\ll J.
\end{array}\right.
\end{equation}
At short times $C_{T_0S}(t)$ experiences a Gaussian decay on a
timescale $\sqrt{J^2+4x_0^2}/4x_0\sigma_0$, while in the case of
strong coupling $J\gg{\rm max}\{X_0,\sigma_0\}$ at long times $t\gg 
J/4\sigma_0^2$ a power law decay appears, \cite{Coish-PRB05}
\begin{equation}
C_{T_0S}(t)\sim C_{T_0S}(\infty)-
\frac{e^{-x_0^2/2\sigma_0^2}}{4\sigma_0\sqrt{J}t} 
\cos\left(Jt+\frac{3\pi}{4}\right).
\end{equation}
Those results show that the singlet-triplet correlator decays due to
the quantum distribution of the nuclear spin system, even for a static
system. For non zero exchange interaction $J\ne 0$ the asymptotic
behavior of the correlator $C_{T_0S}(t)$ changes from a short time
Gaussian behavior  to a long time power-law ($\sim 1/t^{3/2}$) decay
and acquires a universal phase shift which is $3\pi/2$, consistent
with experimental findings for the correlator $C_{SS}(t)$
\cite{Laird-PRL06}. Qualitatively similar results appear when looking
at the transverse correlator in the $S_z=0$ subspace, though one finds 
different decay power and different value of the universal phase
shift.

\subsection{Nuclear spin state manipulation}

As mentioned in the previous sections, for a system of $N$ unpolarized
nuclei and an effective hyperfine interaction strength $A$, the
dephasing time in a weak magnetic field is $T_2^*\sim
1/\sigma\sim\sqrt{N}/A$, where $\sigma$ is the width of the
distribution of the nuclear field $h_z$. This decay $T_2^*$ finds its
origin in the ensemble average over the field distribution. In order
to prolong the electron spin coherence, narrowing of the nuclear field
distribution was proposed in \cite{Coish-PRB04} as an alternative to
the strategy of polarizing the nuclear spins \cite{Burkard-PRB99},
that would require a polarization close to 100\% to be efficient,
which is currently not available \cite{Coish-PRB04}. Few methods for
nuclear spin state narrowing have been studied, in
Ref.~\cite{Klauser-PRB06} the narrowing is due to gate-controlled Rabi
oscillations in a double quantum dot in which the exchange interaction
oscillates, in Ref.~\cite{Giedke-PRA06} a scheme based on quantum
phase estimation is envisioned for a single undriven spin in a single
quantum dot, and in Ref.~\cite{Stepanenko-PRL06} the narrowing is
achieved by optical preparation. 

\subsubsection{Nuclear state narrowing by qubit state measurement}

Here we discuss a nuclear state narrowing technique that has been
proposed in \cite{Klauser-PRB06}. Consider for definiteness the ESR 
Hamiltonian Eq.~(\ref{Eq:ESRham}). The effective Zeeman splitting is
given by $b+h^n_z$, where $b=g^*\mu_B B_z$ and $h^n_z$ is an
eigenvalue of $h_z$.
The idea behind state narrowing is that the ESR driving give rise to
the resonance condition $b+h^n_z-\omega=0$, such that the evolution of
the electron spin depends on the nuclear spin state and thus a
determination of the electron spin evolution results in a
determination of the nuclear spin state.  

The eigenvalues of the nuclear field, as already mentioned in the
previous sections, are Gaussian distributed in equilibrium. The
diagonal elements of the nuclear spin density matrix are
$\rho_I(h^n_z,t=0)=\langle h^n_z|\rho_I|h^n_z\rangle=\exp(-(h^n_z-\langle 
h_z\rangle)^2/2\sigma^2)/\sqrt{2\pi}\sigma$, with mean $\langle
h_z\rangle$ and variance $\sigma$. Therefore, initializing the
electron spin in the state $|\uparrow\rangle$ at time $t=0$, the
probability to find the electron spin in the state
$|\downarrow\rangle$ is given by 
\begin{equation}
P_{\downarrow}(t)=\int dh_z^n\rho_i(h^n_z,0)P^n_{\downarrow}(t),
\end{equation}
where $P^n_{\downarrow}(t)$ is the probability to find the electron
spin in the state $|\downarrow\rangle$, for a given an eigenvalue
$h_z^n$ of the nuclear field $h_z$,  
\begin{eqnarray}
P^n_{\downarrow}(t)&=&\left|\langle h^n_z|\otimes\langle\downarrow| 
U^{\rm ESR}(t)|\uparrow\rangle\otimes|h^n_z\rangle\right|^2\nonumber\\
&=&\frac{1}{2}\frac{b_{\rm ac}^2}{b_{\rm ac}^2+4\delta_n^2}
\left[1-\cos\left(\frac{t}{2}\sqrt{b^2_{\rm ac}+4\delta^2_n}
\right)\right].
\end{eqnarray}
If at time $t=t_m$ we perform a measurement of the electron spin and
find $|\downarrow\rangle$, the diagonal element of the nuclear spin
density matrix will change according to
\begin{equation}
\rho_I(h^n_z,0)\rightarrow\rho_I^{(1,\downarrow)}(h^n_z,t_m)=
\rho_I(h^n_z,0)\frac{P^n_{\downarrow}(t_m)}{P_{\downarrow}(t_m)}.
\end{equation} 
In the case where a measurement is performed with a low time
resolution $\Delta t$, $\Delta t\gg 1/b$, such that it gives the time
averaged value, the probability turns out to be $P^n_{\downarrow}=
\lim_{T\rightarrow\infty}(1/T)\int^T_0dtP^n_{\downarrow}(t)=
b^2_{\rm ac}/2(b^2_{\rm ac}+4\delta^2_n)$. Therefore, a measurement on
the electron spin with outcome  $|\downarrow\rangle$ results in a
multiplication of the nuclear spin density matrix by a Lorentzian, with
width $b_{\rm ac}$, centered around the $h^n_z$ that satisfied the condition
$b+h_z^n-\omega=0$. The nuclear spin distribution, thus, undergoes a
narrowing, resulting in an enhancement of the electron spin coherence,
if $b_{\rm ac}<\sigma$. In the case that the measurement outcome is
$|\uparrow\rangle$ the diagonal element of the nuclear spin density
matrix will change according to  
\begin{equation}
\rho_I(h^n_z,0)\rightarrow\rho_I^{(1,\uparrow)}(h^n_z,t_m)=
\rho_I(h^n_z,0)\frac{1-P^n_{\downarrow}(t_m)}{1-P_{\downarrow}(t_m)},
\end{equation} 
resulting in a reduced probability for the nuclear field to have a
value that matches the resonance condition $b+h_z^n-\omega=0$. 

This procedure can be iterated many times before changes due to the
slow internal dynamics start to affect the nuclear spin state. Many
measurement of the electron spin are possible within this time, with
re-initialization of the electron spin state between the
measurements. Assuming that $M$ cycles can be performed with a static
nuclear field, we have   
\begin{equation}
\rho_I(h^n_z,0)\rightarrow\rho_I^{(M,\alpha_{\uparrow})}(h^n_z,t_m)=\frac{1}{N}
\rho_I(h^n_z,0)(P^n_{\downarrow})^{\alpha_{\uparrow}}
(1-P_{\downarrow})^{M-\alpha_{\uparrow}},
\end{equation} 
where $\alpha_{\uparrow}$ is the number of measurement outcomes
$|\downarrow\rangle$. If the outcome is $|\downarrow\rangle$ the
narrowing has been achieved, otherwise, it is necessary to wait for a
re-equilibration of the nuclear system before the next measurement.

\subsubsection{Optical preparation of nuclear spins}

\begin{figure}
 \begin{center}
 \includegraphics[width=6cm]{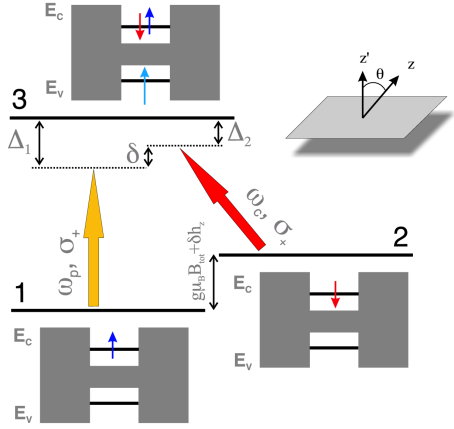}
    \caption{Three-level system. The states 1(2) are spin-up (-down)
      conduction-band ($E_C$) electron, with splitting given by
      $g\mu_BB_{\rm tot}+\delta h_z$, where $\delta h_z$ is the $z$
      component of the nuclear field fluctuations. State 3 is a trion
      with $J_{z'}=3/2$ (With permission from the Authors).\label{Fig:Trion}} 
 \end{center}
\end{figure}

Here we discuss the case of optical nuclear spin preparation that 
makes use of spin-flip two-photon Raman resonance in a driven three-level 
system (TLS) \cite{Stepanenko-PRL06}. 
The lowest electronic states in GaAs quantum dots that are optically 
active under $\sigma_+$ circularly polarized excitation are the ground
state of a single localized conduction-band ($E_C$) electron, in which a
Zeeman field splits the up and down spin states, and the negatively
charged exciton (trion) $|X\rangle$, given by two electrons with
antiparallel spin plus one valence band  heavy hole (hh) with angular
momentum $J_{z'}=+3/2$, as schematically shown in
Fig.~\ref{Fig:Trion}.  The $J=3/2$ subspace in the valence band split
up into heavy and light holes (hh and lh) along the direction $z'$ of
strong quantum dot confinement, that is in general different from the 
$z$-axis in the conduction band, chosen to be the direction of the
magnetic field $B$. The two 
circularly polarized lasers stimulate the transition between
$|\uparrow\rangle$ and $|X\rangle$ at frequency
$\omega_p=\omega_X-\omega_{\uparrow}-\Delta_1$ and the transition between
$|\downarrow\rangle$ and $|X\rangle$ at frequency
$\omega_c=\omega_X-\omega_{\downarrow}-\Delta_2$, while the trion
$J_{z'}=-3/2$ is not excited.   

The narrowing of the nuclear field distribution is based on light
scattering in the TLS, where two long-lived spin states are resonantly 
coupled to the excited trion state $|X\rangle$ that decays
spontaneously. For the two-photon resonance condition
$\delta=\Delta_1-\Delta_2=0$, where $\delta$ is the detuning of the
difference of the frequency of the two lasers $\omega_c-\omega_p$ from
the Zeeman splitting $\omega_z$ of the two spin states, the system is
in a superposition of the two spin 
states with a vanishing excited state component, and the system is 
driven to a dark state with no photon emission. In presence of a nuclear 
spin field, the resonance moves to $\delta=\delta h_z$, where 
$\delta h_z$ is the deviation of the Overhauser field from its mean. 
Monitoring the photon emission constitutes a continuous weak measurement 
of the Overhauser field $h_z$. The absence of photon emission in the
limit $t\rightarrow\infty$, corresponding to the strong measurement
limit, would project the nuclear state onto $|\delta h_z=0\rangle$,
with width $\sigma=0$, therefore letting the dephasing time to diverge,  
$T_2^*\sim 1/\sigma\rightarrow\infty$. A continuous weak measurement
of the Overhauser field, supported by an adaptive adjustment of the
lasers frequencies every time a photon is detected, leads to a
narrowing of the nuclear field distribution, and an enhancement of the
phase coherence of the electron spin.

The relevant effective Hamiltonian of the TLS in the rotating wave
approximation is block diagonal, with blocks labeled by the
eigenvalues $\delta h^k_z$ of the field $\delta h_z$
\begin{equation}
{\cal H}_k=-\frac{\hbar}{2}
\left(\begin{array}{ccc}
\delta h^k_z+\delta & 0 & \Omega_p\\
0 & -\delta h^k_z-\delta & \Omega_c\\
\Omega_p & \Omega_c & -\Delta
\end{array}\right),
\end{equation}
where $\Delta=\Delta_1+\Delta_2$. The combined system consisting of
the TLS and the nuclear spins evolves in time according to a
generalized master equation $\dot{\rho}={\cal L}\rho$, where ${\cal
  L}$ is the Liouvillian operator defined as
\begin{equation}
\dot{\rho}={\cal L}\rho\equiv\frac{1}{i\hbar}[H,\rho]+{\cal W}\rho,
\end{equation}
where $H$ is the Hamiltonian of the system, and ${\cal
  W}=\sum_{\alpha=\uparrow,\downarrow}\Gamma_{X\alpha}(2\sigma_{\alpha
  X}\rho\sigma_{X\alpha}-\sigma_{XX}\rho-\rho\sigma_{XX})/2+
\sum_{\beta=\downarrow,X}\gamma_{\beta}(2\sigma_{\beta\beta}\rho\sigma_{\beta\beta}
-\sigma_{\beta\beta}\rho-\rho\sigma_{\beta\beta})/2$. Here the rate
$\Gamma_{X\alpha}$ describes the radiative decay of $|X\rangle$ into
$\alpha=|\uparrow\rangle$, $|\downarrow\rangle$, while
$\gamma_{\beta}$ is the pure dephasing rate of state
$\beta=|\downarrow\rangle$, $|X\rangle$ with respect to
$|\uparrow\rangle$.

Taking as initial state a product
of arbitrary density matrices $\chi_0$ for the TLS and
$\nu_0=\sum_{kk'}\nu_{kk'}|\delta h^k_z\rangle\langle h^{k'}_z|$ for
the nuclear field, the stationary solution is an entangled state
$\bar{\rho}$.  

In order to describe the
state of the system conditional on a measurement record, a conditional
density matrix is used. The {\it a posteriori} distribution $\nu_{kk}$
is found to be concentrated around the two photon resonance. 
The stationary emission rate is 
\begin{equation}
\Gamma_{\rm em}={\rm Tr}{\cal S}\bar{\rho}(t)=
\Gamma\sum_k\nu_{kk}\langle X|\rho_{kk}|X\rangle,
\end{equation}
where $\Gamma=\Gamma_{X\uparrow}+\Gamma_{X\downarrow}$ and ${\cal S}$ 
is the collapse operator, describing spontaneous emission of the
state $|X\rangle$ into $|\uparrow\rangle$ and $|\downarrow\rangle$ at
rates $\Gamma_{X\uparrow}$ and $\Gamma_{X\downarrow}$. the update rule
for $\nu$ upon photon emission is
\begin{equation}
\nu'_{kk}=\frac{\nu_{kk}\langle X|\rho_{kk}|X\rangle}
{\sum_j\nu_{jj}\langle X|\rho_{jj}|X\rangle}.
\end{equation}
The population in the Overhauser field $\delta h_z$ corresponding to
the two-photon resonance $\delta h_z=\delta$ is depleted by the photon
emission. The electron spin coherence is quantified by the time
dependence of the transverse electron spin component, which in turn is
given by the Fourier transform of the nuclear field distribution,
$\langle S_+(t)\rangle=(\hbar/2)\sum_k\nu_{kk}\exp(it\delta
h^k_z)$. The repeated observation of the quantum dot photon emission
and consequent adaption of the laser frequencies after each photon
detection leads to a narrowing in the nuclear distribution and
consequent enhancement of the electron spin coherence time.

\subsubsection{Exponential decay in narrowed nuclear state}

We have seen that the nuclear spin bath induces a non-Markovian
dynamics of the electron spin, with super-exponential or power-law
decay of the correlation functions. On the other hand it has been
argued that a narrowing of the nuclear spin distribution is expected
to prolong the electron spin coherence. 
In \cite{Coish-arxiv07} it is shown that, in case of a large Zeeman
splitting and for a particular narrowed nuclear spin state, a
Markovian dynamics can arise from virtual flip-flops between the
electron spin and the nuclear spin system, with simple exponential
decay. 

The Hamiltonian describing the interaction of the electron spin with
the nuclear system in a large magnetic field is given by
Eq.~(\ref{Eq:HamiltLargeZeeman}). The energy non-conserving term $V$
can be eliminated by means of a Schrieffer-Wolff transformation, 
$\bar{\cal H}=e^S{\cal H}e^{-S}\approx H={\cal H}_0+\frac{1}{2}[S,V]$, 
where $S={\cal L}_0V$, and ${\cal L}_0$ is the unperturbed
Liouvillian, defined by ${\cal L}_0O=[{\cal H}_0,O]$.  The effective
Hamiltonian $H$ is given by
\begin{equation}
H=(\omega+X)S^z+D.
\end{equation} 
The operators $\omega$, $D$ and $X$ are nuclear spin operators and the
first two are diagonal in a product-state basis
of $I^z_k$-eigenstates, whereas $X$ is purely off-diagonal and
produces correlations between nuclear spins. Corrections of
the order of $\sim A^2/Nb$ in the diagonal terms of $H$ are neglected,
whereas the term of this size in $X$ are retained. This
assumption is valid as long as the bath correlation time $\tau_c$ is
much shorter than the time scale after which the diagonal corrections
become relevant for $b\gg A$, where a Born-Markov approximation is
valid: $\tau_c\sim N/A\ll Nb/A^2$. As a result $\omega=b+h^z$.

The electron and nuclear states are assumed to be initially unentangled
and the nuclear system is prepared in a narrowed state,
$\omega|n\rangle=\omega_n|n\rangle$. For this initial conditions, the
dynamics of the transverse electron spin  component $\langle
S_+\rangle_t$ is described by a GME, and can be written in a rotating
frame defined by $x_t=\exp[-i(\omega_n+\Delta\omega)t]\langle
S_+\rangle_t$, where $\Delta\omega$ is a frequency shift
self-consistently defined by $\Delta\omega=-{\rm Re}\int_0^{\infty}dt
\Sigma(t)$, with
$\tilde{\Sigma}(t)=\exp[-i(\omega_n+\Delta\omega)t]\Sigma(t)$, through
the memory kernel $\Sigma(t)$ of the GME. The equation of motion for
$x_t$ is given by
\begin{equation}
\dot{x}_t=-i\int_0^td\tau\tilde{\Sigma}(\tau)x_{t-\tau}.
\end{equation}    
If $\Sigma(t)$ decays to zero sufficiently fast on the time scale
$\tau_c\ll T_2$, where $T_2$ is in turn the decay time of $x_t$, it is
possible to approximate $x_{t=\tau}\approx x_t$ and extend the upper
limit of the integral to infinity, $t\rightarrow\infty$, obtaining a
Markovian dynamics
\begin{equation}
x_t=\exp(=t/T_2)x_0+\epsilon(t),\qquad \frac{1}{T_2}=-{\rm
  Im}\int_0^{\infty}dt\tilde{\Sigma}(t),
\end{equation} 
where $\epsilon(t)$ gives a small non-Markovian correction that cab be
bounded precisely if $\tilde{\Sigma}(t)$ is known. 

For a homo nuclear system, by expanding $\Sigma(t)$ in the perturbation
$V=XS^z$ and retaining only leading orders in the Born approximation
in the small parameter $A/\omega_n$,
the decoherence time $T_2$ can be cast in the compact form
\begin{equation}\label{Eq:T2expodecayLargeZ}
\frac{1}{T_2}={\rm Re}\int_0^{\infty}dte^{-i\Delta\omega t}\langle
X(t)X\rangle, \qquad X(t)=e^{-i\omega t}Xe^{i\omega t},
\end{equation} 
where the average stands for an expectation value taken with respect
to the initial nuclear state. Though the compact form resembles the
standard result for pure dephasing valid in a weak coupling expansion,
here there is no such weak coupling expansion. The decoherence rate
$1/T_2$ depends on the correlator $C(t)=\langle X(t)X\rangle$. For an
isotropic electron wave function of the form
$\psi(r)=\psi(0)e^{-(r/r_0)^q/2}$ containing $N\gg 1$ nuclei within a
radius $r_0$ in $d$ dimension, the asymptotic dependence of $C(t)$ at
long times is $C(t)\propto 1/t^{2d/q}$, for $t\gg N/A$ and
$d/q<2$. For $d/q\le 1/2$, $1/T_2$ given by
Eq.~(\ref{Eq:T2expodecayLargeZ}) diverges and no Markov
approximation is valid within the Born approximation. On the other
hand, for a 2D dot with a Gaussian electron wave function and for
unpolarized nuclear system, Eq.~(\ref{Eq:T2expodecayLargeZ}) gives the
simple result
\begin{equation}
\frac{1}{T_2}=\frac{\pi}{3}\left(\frac{I(I+1)A}{3b}\right)^2\frac{A}{N}.
\end{equation}
The condition for the validity of the the Markov approximation,
$T_2>\tau_c\sim N/A$ is satisfied whenever $A/b<1$, which correspond
to the range of validity of the Born approximation. Remarkably,
from last equation it follows that $1/T_2$ strongly depends on the
magnitude of the nuclear spin, $1/T_2\propto I^4$. Therefore, systems
with large nuclear spin, such as In ($I_{\rm In}=9/2$), will show
faster decay. 

With these last results on exponential decay in a spin bath, we
conclude the part on electron spin decoherence induced by the nuclear
spin system  and focus on phonon-mediated relaxation of the electron
spin in quantum dots.

\subsection{Phonon-induced relaxation in quantum dots}

Electron spin relaxation in quantum dots takes place via transitions
between spin states, with consequent energy dissipation in the
environment. In a quantum dot the dissipative environment is
represented mainly by the phonons in the surrounding crystal.
Therefore, in order to fully understand relaxation and decoherence
mechanisms that occur in quantum dots, it is important to understand
the manner in which the electron spin interacts with phonons.

Spin-orbit interaction creates an admixture of orbital and spin
degrees of freedom of the electron, and represents an effective coupling
mechanism that mediates the spin-phonon interaction, and that,
ultimately, is responsible for relaxation of the
electron spin. Phonons can produce electric field fluctuations that
can lead to spin relaxation of eigenstates of the spin-orbit
Hamiltonian. Two kinds of electron-phonon interactions are taken into
account, that arise, respectively, from an inhomogeneous deformation of
the crystal potential, resulting in an alteration of the band-gap, and
a homogeneous strain due to piezo-electric effect, the former taking
place in all semiconductors, the latter only in crystals without
structure inversion symmetry such as GaAs.

\subsection{Introduction: Spin-orbit interaction}

An electron that moves in an electric field experiences an effective
magnetic field in its rest frame which interacts with the spin of the
electron. The internal magnetic field depends on the orbital the
electron occupies and therefore spin and orbit are coupled. This well
known effect comes directly from the relativistic Dirac theory of point
particles and it goes under the name of spin-orbit (SO)
interaction. The SO Hamiltonian has the general form
\cite{Sakurai-AQM} 
\begin{equation}
{\cal H}_{\rm SO}=\frac{\hbar}{4m_0^2c^2}{\bf p}\cdot
(\boldsymbol{\sigma}\times{\bf \nabla}V),
\end{equation}  
where $m_0$ is the free electron mass, $c$ is the speed of light,
$\boldsymbol{\sigma}=(\sigma_x,\sigma_y,\sigma_z)$ is the Pauli matrix
vector, and $V$ is the electric potential. In presence of an external
magnetic field ${\bf B}={\bf \nabla}\times{\bf A}$, the canonical
momentum ${\bf p}$ is replaced by the kinetic momentum ${\bf P}={\bf
  p}+e{\bf A}$, ${\bf A}$ being the vector potential. 

In semiconductors like Si or Ge the crystal lattice has
spatial inversion symmetry. For such materials, states of a given
momentum ${\bf k}$ are 4-fold degenerate at $B=0$. In fact due to
time reversal symmetry, $\epsilon_{{\bf k},\uparrow}=\epsilon_{-{\bf
    k},\downarrow}$ holds, and from the inversion symmetry one has
$\epsilon_{{\bf k},\sigma}=\epsilon_{-{\bf k},\sigma}$, such that   
$\epsilon_{{\bf k},\uparrow}=\epsilon_{{\bf k},\downarrow}=$ holds. 

\begin{figure}
 \begin{center}
 \includegraphics[width=8cm]{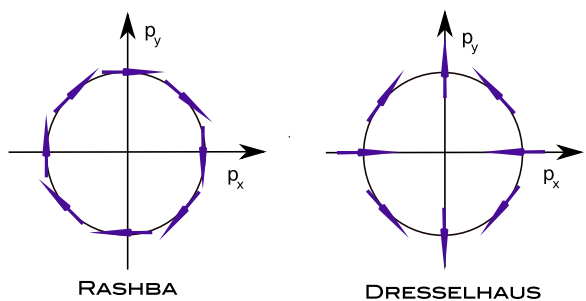}
    \caption{Schematic representation the apparent momentum dependent
      field ${\bf \Omega}({\bf p})$ in the spin-orbit Hamiltonian, 
      ${\bf \Omega}({\bf p})\cdot\boldsymbol{\sigma}$, for the
      Dresselhaus and Rashba spin-orbit interactions.
    \label{Fig:SO-apparent-field}}
 \end{center}
\end{figure}

The double degeneracy can be broken either via the
application of an external magnetic field, which breaks the time
reversal symmetry, or via the brake of spatial inversion
symmetry. This is indeed what happens in crystals that exhibit bulk
inversion asymmetry (BIA), such as the zincblende structure of
GaAs. This effect is know as Dresselhaus spin-orbit interaction
\cite{Dresselhaus-PR55,Dyakonov-SovPhysSem86}. The Hamiltonian for 2D
systems results from the 3D bulk Hamiltonian \cite{Dyakonov-Perel-71} 
after integration over the growth direction $z$ along [001]
\begin{equation}
{\cal H}_D\propto[-\sigma_xp_x\langle p_z^2\rangle+
\sigma_yp_y\langle p_z^2\rangle+\sigma_xp_xp_y^2-
\sigma_yp_yp_x^2]
\end{equation}
where $x$ and $y$ point along the crystallographic directions [100] and
[010]. Due to the strong confinement along $z$, the terms cubic in
momentum components appearing in the Hamiltonian are usually much
smaller than the linear ones, and they are usually
neglected. Retaining only the linear term 
\begin{equation}
{\cal H}_D=\beta(-\sigma_xp_x+\sigma_yp_y)
\end{equation}    
where $\beta$ depends on material properties and on $\langle
p_z^2\rangle$. The spin dynamics resulting from the Dresselhaus
Hamiltonian is well understood in the case of circular orbit, in which
the spin rotates in the opposite direction with respect to the orbit,
as shown in Fig.~\ref{Fig:SO-apparent-field}. 

In heterostructures like GaAs/AlGaAs, an asymmetric confining
potential additionally breaks the inversion symmetry, giving rise a
further spin-orbit interaction due to structural inversion asymmetry
(SIA), known as Bychkov-Rashba term
\cite{Rashba-SovPhysSolSt,Bychkov-JEPT84}. For a confining electric
field along the $z$ direction, the Bychkov-Rashba Hamiltonian
$\propto({\bf E}\times{\bf p})\cdot\boldsymbol{\sigma}$ is
 \begin{equation}
{\cal H}_R=\alpha(\sigma_xp_y-\sigma_yp_x),
\end{equation} 
where $\alpha$ depends on the confining potential and on material
properties. The spin dynamics resulting from the Bychkov-Rashba
Hamiltonian can also be well understood in the case of circular
orbit, in which the spin rotates along in the same direction as the
orbit, being the spin always antiparallel to the direction of motion,
as explained in Fig.~\ref{Fig:SO-apparent-field}.

\subsection{Electron spin relaxation and decoherence}

Due to the fact that the most promising semiconductor devices that
make use of the electron spin as the quantum two level system (qubit)
are realized on the basis of 2D electron gases in GaAs
heterostructures, the following discussion concentrates on spin flip
mechanisms that are relevant for GaAs. 
 
Spin relaxation of localized electrons in quantum dots shows
remarkable differences from the case of delocalized bulk electrons. 
The most effective mechanisms in bulk 2D are related to the broken
inversion symmetry, either the BIA or SIA case, which give rise to a
strong spin-orbit splitting in the electron spectrum, ultimately
responsible for spin flip. Besides, the piezoelectric interaction
arising in non-inversion symmetric crystals provides a strong coupling
of electrons to the bosonic bath of acoustic phonons. The interplay of
these mechanisms results in an efficient spin-lattice relaxation for
bulk carriers in III-V type semiconductors and heterostructures. The
strong localization of electrons in quantum dots leads to suppression
of spin-flip rate. The phonon-assisted spin-flip mechanisms in
semiconductor quantum dots have been studied 
in \cite{Khaetskii-PRB00,Khaetskii-PRB01}.

\subsubsection{Electron spin relaxation in quantum dots}

In the case of strong confinement in the $z$ direction, corresponding to
the [100] crystallographic axis, for a lateral dot size much larger
than the degree of vertical confinement, the Hamiltonian derived from
the Kane model \cite{Pikus-84} for 2D electrons in the conduction band
in the presence of an external magnetic field ${\bf B}$ is
\cite{Khaetskii-PRB00,Khaetskii-PRB01}
\begin{eqnarray}\label{Eq:SODotHam}
{\cal H}&=&\frac{{\bf p}^2}{2m}+U({\bf r})+U_{\rm ph}({\bf r},t)
+\frac{1}{2}g\mu_B\boldsymbol{\sigma}\cdot{\bf B}+{\cal
  H}_{SO}^D+{\cal H}_{SO}^R,\\
{\cal H}_{SO}^D&=&\beta(-\sigma_xp_x+\sigma_yp_y),\qquad 
{\cal H}_{SO}^R=\alpha(\sigma_xp_y-\sigma_yp_x).
\end{eqnarray}
Here ${\bf p}=-i\hbar{\bf \nabla}+(e/c){\bf A}$ is the kinetic
momentum, $m$ the effective mass, $g$ the effective electron
$g$-factor (in GaAs $g=-0.44$), and $\boldsymbol{\sigma}$ the Pauli
matrix vector. The axes $x$, $y$, and $z$ coincide with the main
crystallographic ones, with $z$ perpendicular to the 2D plane. 
The first two terms of the Hamiltonian describe the quantum dot with
confining potential $U({\bf r})$, that is typically chosen
parabolic. The third term describes the spin-independent interaction
with acoustic phonon. The fourth term is the Zeeman
Hamiltonian. ${\cal H}_{SO}$ describes the spin-orbit effects. ${\cal
  H}_{SO}^D$ is the Dresselhaus term, due to BIA, and 
${\cal H}_{SO}^D$ is the Rashba term, due to SIA. For GaAs
heterostructures $\beta\approx 10^5$ cm/s. 

The Hamiltonian of Eq.~(\ref{Eq:SODotHam}) should contain also term
describing ``direct'' interaction between spin and phonons, such as
that due to an inhomogeneous deformation of the lattice, and a term
describing the spin-phonon coupling in presence of a magnetic field
due to a lattice-deformation-dependent admixture of valence-band and 
conduction-band states. Their contribution on spin relaxation rates
turns out to be negligible with respect to the dominant admixture
mechanism contribution ascribable to the Dresselhaus and Rashba
spin-orbit interaction.  See Ref.~\cite{Khaetskii-PRB00,
Khaetskii-PRB01,Roth-PR60} for a discussion of the direct spin-phonon
coupling contribution. 

The phonon-induced rate for the transition between
$|\Psi_n^{\uparrow}\rangle$ and $|\Psi_n^{\downarrow}\rangle$ is given
by Fermi's golden rule 
\begin{equation}
\Gamma=\frac{2\pi}{\hbar}\sum_n|\langle\Psi_n^{\uparrow}|{\cal
  H}_{ph}|\Psi_n^{\downarrow}\rangle|^2D(\epsilon_Z). 
\end{equation}
Here $D(E)$ is the phonon density of states at the Zeeman energy
splitting $\epsilon_z$. From
experimental results, the relevant acoustic phonons can be treated as
bulk-like phonons, showing a linear dispersion relation in the
relevant energy range, for which the density of states increases
quadratically with energy \cite{Fujisawa-Science98}.
The states $|\Psi_n^{\uparrow}\rangle$ and
$|\Psi_n^{\downarrow}\rangle$ are the effective spin states,
containing more than one orbital and both the spin {\it up} and {\it
  down} states. This admixture of spin and orbit comes out in taking
into account the spin-orbit interaction due to BIA and SIA as a
perturbation. Due to the localization of stationary states in a quantum
dot, it follows that the spin-orbit interaction does not directly
couple Zeeman-split sublevels in the same quantum dot orbital. It
follows that within first order perturbation theory in the spin-orbit
Hamiltonian, the effective single electron quantum dot states are
\begin{eqnarray}
|\Psi_n^{\uparrow}\rangle&=&|n\uparrow\rangle+\sum_{n'\ne n}
\frac{({\cal H}_{SO})_{n'n}^{\downarrow\uparrow}}{E_n-E_{n'}
+g\mu_BB}|n'\downarrow\rangle,\\ 
|\Psi_n^{\downarrow}\rangle&=&|n\downarrow\rangle+\sum_{n'\ne n}
\frac{({\cal H}_{SO})_{n'n}^{\uparrow\downarrow}}{E_n-E_{n'}
-g\mu_BB}|n'\uparrow\rangle, 
\end{eqnarray}    
where $({\cal H}_{SO})_{n'n}^{\downarrow\uparrow}=
\langle n'\downarrow|{\cal H}_{SO}|n\uparrow\rangle$ and 
$\{|n\rangle\}$ are the unperturbed quantum dot orbital states. 
Due to the anisotropy of BIA and SIA spin-orbit interaction,
the admixture of spin and orbit degrees of freedom turns out to be 
anisotropic \cite{Falko-PRL05}.

For spin-flip transitions involving a small energy transfer, the
dominant contribution comes from piezo-electric phonons. 
The electric field associated with a single phonon scales like
$1/\sqrt{q}$ for piezo-phonons and like $\sqrt{q}$ for deformation
potential phonons, $q$ being the phonon wave number. This is due to
the fact that piezo-phonons come from a homogeneous lattice strain, in
which long wavelengths play a major role. Vice versa, a local deformation
would involve short wavelengths, and so higher energies. On the other
hand, wavelengths much longer than the dot size give rise to a global
shift of the entire dot potential, therefore the effective phonon
wavelengths are those comparable with the dot size, as seen in
\cite{Meunier-PRL07}. 
\begin{figure}
 \begin{center}
 \includegraphics[width=6cm]{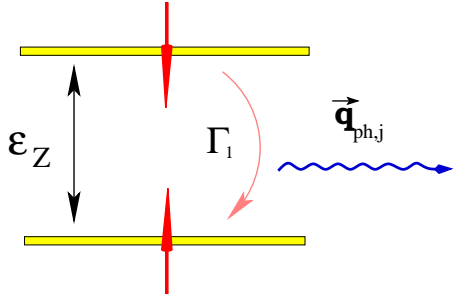}
    \caption{Schematic representation of the spin-flip process
      associated with the emission of a phonon of energy $\epsilon_Z$
      and momentum ${\bf q}_{j}$, where $j$ is the branch index, at
      rate $\Gamma_1$.
    \label{Fig:Relaxation-PhononEmission}}
 \end{center}
\end{figure}

The electron-phonon coupling for piezo-electric phonons has the form
\begin{equation}
U_{ph}^{{\bf q}\alpha}({\bf r},t)\propto\exp(i{\bf q}\cdot{\bf r}
-i\omega_{{\bf q}\alpha}t)A_{{\bf q}\alpha}b^{\dag}_{{\bf q}\alpha}+h.c.,
\end{equation}   
where ${\bf q}\alpha$ are the wavelength and the branch of the
phononic modes, $A_{{\bf q}\alpha}$ is effective anisotropic
piezo-electric modulus of wave ${\bf q}\alpha$. The matrix elements of
the phonon Hamiltonian between the Zeeman split sublevels of orbital
level $n$, that describe the spin-flip process with emission of phonon
${\bf q}\alpha$, are given at first order in the spin-orbit
interaction by 
\begin{equation}\label{Eq:piezo-matr-el}
\langle \Psi_n^{\uparrow}|U_{ph}^{{\bf q}\alpha}|
\Psi_n^{\downarrow}\rangle=\sum_{k\ne n}
\left[\frac{(U_{ph}^{{\bf q}\alpha})_{nk}
({\cal H}_{SO})_{kn}^{\uparrow\downarrow}}{E_n-E_k-g\mu_BB}
+\frac{(U_{ph}^{{\bf q}\alpha})_{nk}
({\cal H}_{SO})_{kn}^{\downarrow\uparrow}}{E_n-E_k+g\mu_BB}\right]. 
\end{equation}
As a consequence of Kramer's theorem, in case of no external magnetic
field, Eq.~(\ref{Eq:piezo-matr-el}) is zero. Considering only 
Dresselhaus spin-orbit interaction, for small Zeeman
splitting, $g\mu_BB\ll \sqrt{ms^2\hbar\omega_0}$, it is possible to
obtain an effective spin-flip Hamiltonian which acts on the subspace
of Zeeman sublevels of orbital level $n$, \cite{Khaetskii-PRB01},
where a phonon induced electric field arise as a gradient of
$U_{ph}$. For a parabolic dot confinement potential, and for the
particular case of circular dot with level spacing $\omega_0$,
the spin-flip rate for the transition between the Zeeman sublevels
of the dot ground state, associated with the emission of a
piezo-phonon as depicted in Fig.~\ref{Fig:Relaxation-PhononEmission},
is  \cite{Khaetskii-PRB01} 
\begin{equation}\label{Eq:G1-dressel-el}
\Gamma_1=\frac{(g\mu_BB)^5}{\hbar(\hbar\omega_0)^4}
\Lambda_p(1+\cos^2(\vartheta)),\qquad 
\Lambda_p\equiv\frac{2}{35\pi}\frac{(eh_{14})^2\beta^2}{\rho\hbar}
\left(\frac{1}{s_l^5}+\frac{4}{3s_t^5}\right),
\end{equation}  
where $\beta$ is the strength of the Dresselhaus spin-orbit
interaction, $\vartheta$ is the angle between the direction of
confinement in the quantum dot $z$, and the direction of the applied
magnetic field $z'$, and $\Lambda_p$ is the strength of the effective
spin-piezo-phonon coupling. For given longitudinal and transverse
sound speed $s_l$ and $s_t$, crystal mass density $\rho$, and modulus
of the piezo-tensor $eh_{14}$ ($eh_{14}=1.3\times 10^7$ eV/cm for
GaAs), it ranges from $\approx 7\times 10^{-3}$ to $\approx 6\times
10^{-2}$, depending on $\beta$. Eq.~(\ref{Eq:G1-dressel-el}) shows a
strong dependence on the magnetic field and the lateral dot
confinement energy $\omega_0$. For $\hbar\omega_0=10$ K and magnetic
field $B=1$ T, $\Gamma_1\approx 1.5\times 10^3{\rm s}^{-1}$. These
theoretical expectations, in particular the $B$-dependence, have been
confirmed in experiments 
\cite{Rugar-Nature04,Elzerman-Nature04,Kroutvar-Nature04}, and
long spin relaxation time, up to $1{\rm s}$, have been measured
\cite{Amasha-arXiv07}.

The effect of the Rashba spin-orbit interaction has so far not been
taken into account. As the Dresselhaus term, it contributes to the
admixing of spin and orbital states, and therefore to relaxation due
to phonon scattering. The effect of the interplay of these two terms
can show up in a strong difference in the their associated relaxation
rates \cite{Amasha-arXiv07,Bulaev-PRB05,Stano-PRB05,Stano-PRL06}. 
For a quantum dot in external magnetic field, the first and
second lowest levels show a crossing behavior as a function of the
applied magnetic field, the ground state being not affected. In a
perturbative treatment of the spin-orbit interaction
\cite{Bulaev-PRB05}, Dresselhaus and
Rashba terms show a qualitative difference, in which the latter
couples the crossing levels, giving rise to an anticrossing of the
levels at the point of accidental degeneracy. For magnetic fields
much smaller than the crossing level value, the these two levels have
a well defined spin orientation, i.e. low degree of admixture. In the
region of anticrossing the admixture leads to complete superposition of
the {\it up} and {\it down} states, and eventually to a reversed
situation in the limit of magnetic field much larger than the crossing
value, in which the two levels have again well defined spin, but
reversed. Therefore, sweeping the magnetic field over the crossing
region leads to spin-flip. In particular at the avoided crossing
point, the strong admixture between spin states lead to a cusp-like
behavior of the relaxation rate as a function of the magnetic field,
at the anticrossing point.         

\subsubsection{Phonon-induced electron spin decoherence}

\begin{figure}
 \begin{center}
 \includegraphics[width=7cm]{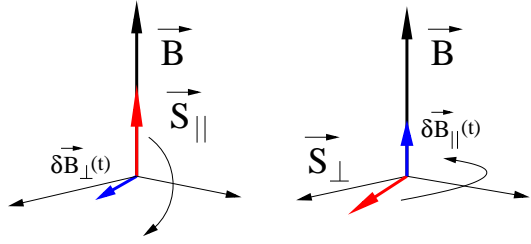}
    \caption{Schematic representation of the relaxation process (left
      side), due to magnetic field fluctuations $\delta B_{\perp}(t)$
      orthogonal to the applied $B$ field, and of the dephasing
      process (right side), due to magnetic field fluctuations $\delta
      B_{\parallel}(t)$ parallel to the applied $B$ field.
    \label{Fig:ParallelVsTransverse}}
 \end{center}
\end{figure}
In a Markovian dynamics the decoherence time $T_2$ is limited by both
spin-flip and dephasing processes, though its upper bound is $T_2\le
2T_1$.   
A systematic analysis of phonon-induced spin decay
is carried out in \cite{Golovach-PRL04}. Also there both the Rashba and
Dresselhaus spin-orbit interactions are treated perturbatively.
The deformation potential phonons are also considered. For a spin-orbit
length $\lambda_{SO}=\hbar/m^*\beta$ much larger than the electron
orbit size $\lambda$, the contribution to the spin-phonon coupling in
the Hamiltonian 
linear in $\lambda/\lambda_{SO}\propto\alpha,\beta$ is due only to a
finite Zeeman splitting. For $B$ field in the range $m^*\beta^2\ll
g\mu_BB\ll\hbar\omega_0$, the effective Hamiltonian is
\begin{equation}\label{Eq:SO-Eff-Ham}
{\cal H}_{\rm eff}=\frac{1}{2}g\mu_B[{\bf B}+\delta{\bf
  B}(t)]\cdot\boldsymbol{\sigma},\qquad
\delta{\bf B}(t)=2{\bf B}\times{\bf \Omega}(t), 
\end{equation} 
where ${\bf \Omega}(t)=\langle\psi|[(\hat{L}_d^{-1}\boldsymbol{\xi}), 
U_{\rm ph}(t)]|\psi\rangle$, $|\psi\rangle$ is the electron orbital
wave function, $\hat{L}_d$ is the dot Liouvillian, $\hat{L}_dA=[{\cal
  H}_d,A]$. The vector $\boldsymbol{\xi}$ lies in the 2D dot plane and
depends on $\alpha$, $\beta$, and $m^*$. The most important
consequence of Eq.~(\ref{Eq:SO-Eff-Ham}) is that within first order in the
spin-orbit coupling parameter there can only be transverse
fluctuations of the effective field.   
 
In case of many uncorrelated scattering events, the expectation values
$\langle{\bf S}\rangle$ of the spin obeys the Bloch equation
\cite{Slichter} 
\begin{equation}\label{Eq:Bloch-phonon}
\langle\dot{\bf S}\rangle=\boldsymbol{\omega}\times
\langle{\bf S}\rangle-\Gamma\langle{\bf S}\rangle+
{\bf \Upsilon},
\end{equation}
where $\boldsymbol{\omega}=\omega{\bf B}/B$, $\omega=g\mu_BB/\hbar$. 
In the Born-Markov approximation, for a
generic $\delta{\bf B}$ such that $\langle\delta{\bf B}\rangle=0$, the
tensor $\Gamma$ can be expressed in terms of the spectral function
\begin{equation}
J_{ij}(w)=\frac{g^2\mu_B^2}{2\hbar^2}\int_0^{\infty}dt
\langle\delta B_i(0)\delta B_j(t)\rangle e^{-iwt},
\end{equation}
and it is diagonal in a frame $(X,Y,Z)$, with $Z$ oriented along ${\bf
B}$. The symmetric part of $\Gamma$ is responsible for the decay of
the spin components, and it can be expressed just as function of
$J^{\pm}_{ij}(w)={\rm Re}[J_{ij}(w)\pm J_{ij}(-w)]$. $\Gamma$ can be
split in two contributions, $\Gamma=\Gamma^r+\Gamma^d$, where
$\Gamma^r$ contains the spectral function $J_{ij}(\omega)$ at the
Zeeman frequency $\omega$, and describes spin decay due to emission or
absorption of a phonon, whereas $\Gamma^d$ is due to elastic
scattering of spin. The 
relaxation time $T_1$ is completely determined by $\Gamma^r$, while
the decoherence time $T_2$ is affected by both  $\Gamma^r$ and
$\Gamma^d$, the latter describing pure dephasing. 
In the general case, solution of Eq.~(\ref{Eq:Bloch-phonon}) yields
\begin{equation}
\frac{1}{T_1}:=\Gamma_{ZZ}=\Gamma_{ZZ}^r,\qquad
\frac{1}{T_2}:=\frac{1}{2}(\Gamma_{XX}+\Gamma_{YY}).
\end{equation}
In many cases the contribution of $\Gamma^r$ to spin decoherence is
negligible, the 
decoherence rate being determined entirely by $\Gamma^d$. However, it
turns out that, at first order perturbation theory in the spin-orbit
interaction, no dephasing takes place \cite{Golovach-PRL04}. Due to
the transverse nature of the fluctuations in the magnetic field in the
effective Hamiltonian, the tensor $\Gamma^d$ is identically zero,
$\Gamma^d=0$ (Fig.~\ref{Fig:ParallelVsTransverse} illustrates an
intuitive picture of the 
effect of the longitudinal and the transverse fluctuations). As a result 
\begin{equation}
\frac{1}{T_1}=\frac{2}{T_2}=J^+_{XX}(\omega)+J^+_{YY}(\omega).
\end{equation}    
Contributions to the decoherence time $T_2$ due to pure dephasing
arise when two-phonon processes are taken into account in the next
order in the electron-phonon interaction \cite{SanJose-PRL06}. 
Therefore, if only 
spin-orbit decay mechanisms are taken into account, the decoherence
time $T_2$ for the decay of the transverse component of an electron
spin in GaAs quantum dots is $T_2=2T_1$.

\subsection{Spin-orbit interaction for heavy holes}

The electron spin in GaAs quantum dots has shown to have a long
relaxation time, due to inefficient phonon-induced relaxation
mechanisms. On the other hand, the decoherence time is mainly
dominated by hyperfine induced decay, due to the fact that the decay of
the longitudinal electron spin component can be strongly suppressed
by the application of an external magnetic field.  
In order to circumvent this problem, the use of hole spins as carriers
has been recently proposed. The valence band in III-V semiconductors
has a $p$ symmetry, for which the electron has zero probability to be
found on the position of the nucleus. According to
Eq.~(\ref{Eq:HyHam}), it follows that the hyperfine interaction
between holes and nuclei is strongly suppressed with respect to that
of nuclei and conduction band electrons.        
However, the hole spin relaxation time turns out to be much smaller than
that of the electrons by several order of magnitude. The reason for
this is due to the fact that, beside the spin-orbit coupling due to
bulk inversion asymmetry and the structural inversion asymmetry, there
is a strong spin-orbit coupling between the heavy-holes (HH) and the
light-holes (LH) sub-bands \cite{Efros-PRB98}. Investigations of hole
spin relaxation in quantum dots, exclusively due to spin-orbit
coupling of LH and HH sub-bands, give estimates for the relaxation
time much shorter than the case for electron spin
\cite{Woods-PRB04,Lu-PRB05}.  

In Ref.~\cite{Bulaev-PRL05}, HH spin relaxation is analyzed in presence of
Rashba and Dresselhaus spin-orbit coupling, as well as spin-orbit
between HH and LH. From the two-band Kane model, the Hamiltonian for
the valence band of III-V semiconductors is given by
\begin{equation}\label{Eq:Hole-bulkHam}
{\cal H}_{\rm bulk}={\cal H}_{\rm LK}+\eta{\bf
  J}\cdot{\bf \Omega}+{\cal H}_{\rm Z},
\end{equation}
where ${\cal H}_{\rm LK}$ is the Luttinger-Kohn Hamiltonian
\cite{Luttinger-PR55}, 
$\eta\propto(E_{\rm g}+\Delta_{\rm so})/\Delta_{\rm so}$, $\Delta_{\rm so}$
is the split-off gap energy, and $E_{\rm g}$ is the band gap
energy. ${\bf J}=(J_x,J_y,J_z)$ are the $4\times 4$ matrices
corresponding to spin 3/2, $\Omega_z=P_z(P_x^2-P_y^2)$, and
$\Omega_x$, $\Omega_y$ are obtained by cyclic permutations. The last
term in Eq.~(\ref{Eq:Hole-bulkHam}) ${\cal H}_Z=-2\kappa\mu_B{\bf
  B}\cdot{\bf J}-2q\mu_B{\bf B}\cdot{\cal J}$ is the Zeeman term for
the valence band \cite{Luttinger-PR56}, with $\kappa$ and $q$
Luttinger parameters \cite{Luttinger-PR56}, and ${\cal
  J}=(J_X^3,J_y^3,J_z^3)$.   

In case of structure inversion asymmetry along the growth direction,
due to an asymmetric confinement, there is an additional contribution
to the spin-orbit interaction, the Bychkov-Rashba term. For the
two-band Kane model it is given by \cite{Winkler-PRB00,Winkler-PRB02} 
$\alpha_R({\bf P}\times{\bf E})\cdot{\bf J}$, where ${\bf E}$ is the
effective electric field along the growth direction, and $\alpha_R$
is the Bychkov-Rashba spin-orbit coupling constant. We consider a
two-dimensional system grown along the [001]-direction. Because of
confinement, the valence band splits into a heavy-hole subband, with
$J_z=\pm 3/2$, and a light-hole subband, with $J_z=\pm 1/2$
\cite{Efros-PRB98,Bulaev-PRL05}, where $z$ is the growth direction. In
case of large splitting $\Delta$ between HH and LH, the properties of the
two subband s can be described separately, the $J_z=\pm 3/2$ subspace
for HHs and the $J_z=\pm 1/2$  subspace for LHs, using only the
$2\times 2$ submatrices. The HHs submatrices have the properties
$\tilde{J}_x=\tilde{J}_y=0$, and
$\tilde{J}_z-\frac{3}{2}\sigma_z$ \cite{Kesteren-PRB90}. For 
low temperatures only the HH subband is significantly
occupied. Considering only HHs, starting from the bulk Hamiltonian Eq.
(\ref{Eq:Hole-bulkHam}) with the addition of the Bychkov-Rashba term,
at the lowest order in perturbation theory
\cite{Zakharchenya-OpticalOrientation}, it is possible to derive an
effective Hamiltonian  for a quantum dot with lateral confinement
potential $U(x,y)$
\begin{equation}\label{Eq:Ham-HH-dot}
{\cal H}=\frac{1}{2}(P_x^2+P_y^2)+U(x,y)+{\cal H}_{\rm SO}^{\rm
  HH}-\frac{1}{2}g_{zz}\mu_BB_z\sigma_z, 
\end{equation}
where $m$ is the effective HH mass, $g_{zz}$ is the component of the
$g$ factor tensor along the growth direction, and the effect of an
in-plane component of the magnetic field can be neglected due to
strong anisotropy in the HH $g$ factor, $g_{\parallel}\ll g_{zz}$
\cite{Kesteren-PRB90}, as well as the orbital effect of the in-plane
magnetic field, as long as $B_{\parallel}\ll c\hbar/eh^2$, $h$ being
the height of the quantum dot. ${\bf P}={\bf p}+(|e|/c){\bf A}({\bf r})$,
with ${\bf A}({\bf r})=(-yB_z/2,xB_z/2,yB_x-xB_y)$, and 
\begin{equation}\label{Eq:SOI-HH}
{\cal H}_{\rm SO}^{\rm HH}=i\alpha\sigma_+P^3_-+\beta
P_-P_+P_-\sigma_++\gamma B_-P_-^2\sigma_++{\rm h.c.}.
\end{equation}  
The first two terms in the spin-orbit interaction for heavy holes
consist in the Rashba and Dresselhaus contribution, respectively,
while the last term $(\gamma)$ describes the combination of two
effects: the orbital coupling via non-diagonal elements in the
Luttinger-Kohn Hamiltonian $(\propto P_{\pm}^2)$, taken into account
perturbatively, and magnetic coupling via   non-diagonal elements in
the Zeeman term, $(\propto B_{\pm})$ \cite{Luttinger-PR56}. This new
spin-orbit term is unique for heavy holes \cite{Bulaev-PRL07}. In
Eq. (\ref{Eq:SOI-HH}) $\alpha-3\gamma_0\alpha_R\langle
E_z\rangle/2m_0\Delta$, $\beta=-3\gamma_0\eta\langle
P^2_z\rangle/2m_0\Delta$, $\gamma=3\gamma_0\kappa\mu_B/m_0\Delta$,
$\sigma_{\pm}=(\sigma_x\pm\sigma_y)/2$, $P_{\pm}=P_x\pm P_y$, and
$B_{\pm}=B_x\pm B_y$, $m_0$ is the free electron mass, $\gamma_0$ is
the Luttinger parameter \cite{Luttinger-PR56}, $\langle E_z\rangle$ is
the average electric field, and $\Delta$ is the splitting between
heavy and light hole subbands, $\Delta\propto h^{-2}$, where $h$ is the
quantum-dot height. For a quantum dot with characteristic lateral size
$l$, the ratio $\langle {\cal H}_{\rm SO}^{\rm el}\rangle/
\langle {\cal H}_{\rm SO}^{\rm HH}\rangle\propto(l/h)^2$. Therefore for
flat quantum dots, $l/h\gg 1$, the spin-orbit coupling for
heavy holes can be weaker than that for conduction electrons
\cite{Bulaev-PRB05,Golovach-PRL04}. This observation has also been
confirmed experimentally \cite{Heiss-arXiv07}, where the 
spin relaxation rate for heavy holes has shown to be comparable to
that of electrons.    

For vanishing spin-orbit interaction, the spectrum of the Hamiltonian
Eq. (\ref{Eq:Ham-HH-dot}) for a parabolic lateral confinement can be
found through a canonical transformation \cite{Ganichev-PRB04}, and
it is the Fock-Darwin spectrum split by the Zeeman term
\cite{Fock-ZP28,Darwin-Proc30}. In the framework of perturbation
theory \cite{Bulaev-PRB05}, it can be seen that the corrections to the
spectrum due to ${\cal H}_{\rm SO}^{\rm HH}$ arise only at second
order, and the spin-orbit interaction influences the wave functions
more strongly than the energy levels. ${\cal H}_{\rm SO}^{\rm HH}$
couples the two lowest states $|0,\pm 3/2\rangle$ to the states with
opposite spin orientations and different orbital momenta $|l,\mp
3/2\rangle$. The different spin-orbit interactions appearing in
Eq. (\ref{Eq:SOI-HH}) differ by symmetry in the momentum space
\cite{Bulaev-PRB05,Ganichev-PRL04}, and thus produce a mixing of
spin-up and spin-down states, with resulting avoided crossings between
energy levels. We mention here that the levels cross only if
$g_{zz}>0$, therefore in case of GaAs ($g_{zz}>0$) quantum dots an
anticrossing appears, with consequent peak of the relaxation rate as a
function of the magnetic field, at the point where the crossing takes
place, while for InAs ($g_{zz}<0$) quantum dot no crossing and no
cuspic-like behavior of the relaxation rate appear.
The spin-orbit mixing of the heavy-hole states provides
a mechanism of transitions between the states $|0,\pm 3/2\rangle$
through 
emission or absorption of an acoustic phonon, that ultimately
represents the main source of relaxation and decoherence for
heavy-holes \cite{Bulaev-PRL05}.       

Taking into account piezoelectric and deformation potential phonons,
the potential of a phonon with mode ${\bf q}\alpha$ is given by
\cite{Khaetskii-PRB01,Zakharchenya-OpticalOrientation} 
\begin{equation}
U_{{\bf q}\alpha}^{\rm ph}=\sqrt{\frac{\hbar}{2\rho s_{\alpha}qV}}
F(q_z)e^{i{\bf q}_{\parallel}\cdot{\bf r}}\times\left\{wA_{{\bf q}\alpha}
+i\left[\left(a+\frac{b}{2}\right){\bf q}\cdot{\bf d}^{{\bf q}\alpha}
-\frac{3}{2}bq_zd_z^{{\bf q}\alpha}\right]\right\},
\end{equation}  
where ${\bf q}_{\parallel}=(q_x,q_y)$, $a$ and $b$ are constants of
the deformation potential, $V$ the quantum dot volume, $s_{\alpha}$
the sound velocity, $\rho$ the crystal mass, $A_{{\bf q}\alpha}$ the
effective piezoelectric modulus, ${\bf d}_{{\bf q}\alpha}$ the phonon
polarization vector, $F(q_z)$ the form factor, which is determined by
the spread of the electron wave function in the $z$ direction. 

\subsubsection{Spin decoherence and relaxation for heavy holes}

For a single-particle quantum dot, in which an heavy hole can
occupy one of the low-lying levels, some energy levels with same spin
orientation cross, with increasing $B$, the upper Zeeman-split ground
state level. Therefore we consider an $n$-level system, in which the
first $n-1$ levels have same spin orientation, while the $n$-level has
opposite spin. In the context of Bloch-Redfield theory, the Bloch
equations for the spin motion of a heavy hole in such a system are
given, in the interaction picture, by
\begin{eqnarray}
\langle\dot{S}_z\rangle&=&(S_T-\langle S_z\rangle)/T_1-R(t),\\
\langle\dot{S}_x\rangle&=&-\langle S_x\rangle/T_2, 
\langle\dot{S}_y\rangle=-\langle S_y\rangle/T_2,
\end{eqnarray}  
where $R(t)=W_{n1}\rho_{nn}(t)+\sum_{i=1}^{n-1}W_{ni}\rho_{ii}(t)$,
$\rho(t)$ is the density matrix, $W_{ij}$ is the transition rate from
state $j$ to state $i$, $S_T$ is a constant that takes the value
$\langle S_z\rangle$ in the thermodynamic equilibrium $R(t)=0$, and
\begin{equation}
\frac{1}{T_1}=W_{n1}+\sum_{i=1}^{n=1}W_{in},\qquad
\frac{1}{T_2}=\frac{1}{2T_1}+\frac{1}{2}\sum_{i=1}^{n-1}W_{i1},
\end{equation}
where the pure dephasing term, which is due to fluctuations along the
longitudinal $z$ direction, is absent in the decoherence time $T_2$,
because the spectral function is superohmic. The spin motion involves
$n-1$ states and therefore there are $n-1$ transition rates. It can be
shown, by solving the master equation, that for low temperature $\hbar
q_{\alpha}\gg k_BT$, $R(t)\approx 0$ and phonon
absorption is strongly suppressed. In this case only
one relaxation rate contributes to relaxation time $T_1$. In this
limit the last sum in the expression for the decoherence rate can be
neglected, and the decoherence time $T_2$ saturates, $T_2=2T_1$.  
The relaxation rates for the different spin-orbit interactions are
\cite{Bulaev-PRL05} 
\begin{eqnarray}
\frac{1}{T_1^{\rm BR}}&\propto&\alpha^2\omega_z^7\left(
\frac{\omega_+^3}{3\omega_++\omega_Z}-
\frac{\omega_-^3}{3\omega_--\omega_Z}\right)^2,\\
\frac{1}{T_1^{\rm D}}&\propto&\beta^2\omega_z^3\left(
\frac{\omega_+}{\omega_++\omega_Z}-
\frac{\omega_-}{\omega_--\omega_Z}\right)^2,\\
\frac{1}{T_1^{\parallel}}&\propto&\gamma^2B_{\parallel}^2\omega_z^5\left(
\frac{\omega_+^2}{2\omega_++\omega_Z}+
\frac{\omega_-^2}{2\omega_--\omega_Z}\right)^2,
\end{eqnarray}
where $\omega_{\pm}=\Omega\pm\omega_c/2$,
$\Omega=\sqrt{\omega_0^2+\omega^2_c/4}$ $\omega_c=|e|B/mc$ is the
cyclotron frequency, 
$\omega_Z-g_{zz}\mu_BB_z$, and $B_{\parallel}-\sqrt{B_x^2+B_y^2}$. 
In contrast to the case of conduction electrons \cite{Golovach-PRL04},
no interference takes place for heavy holes, and the rates originating
from different spin-orbit terms sum up, giving the total spin
relaxation rate $1/T_1=1/T_1^{\rm BR}+1/T_1^{\rm
  D}+1/T_1^{\parallel}$.   For the case of GaAs quantum dots the
crossing between levels takes place at $\omega_Z=\omega_-$,
$2\omega_-$, and $3\omega_-$, and the strong spin mixing arising causes
cuspic-like peaks in the relaxation rate as a function of the external
field $B$.

\subsubsection{Electric dipole spin resonance for heavy holes}

The possibility of coherent manipulating the spins is of great
importance for spintronics and quantum computation. In case of
conduction electron spin-based electronics, such control is obtained
by the electron spin resonance (ESR). Through the application of short
resonant microwave pulses, arbitrary superpositions of spin-up and
spin-down states can be created. Therefore ESR provides a necessary
tool for single-qubit operations, an essential requirement for quantum
computation. In Rabi oscillations and spin echo experiments
\cite{KoppensNature06}, that are based on this technique, the ESR
signal can be detected by measuring the absorption of radio-frequency
(rf) power \cite{Meisels-SemicondST05}. ESR methods involve magnetic-dipole
transitions induced by oscillating magnetic fields. Besides, an
alternative is provided by alternating electric fields, that give rise
to electric-dipole spin resonance (EDSR). 

Considering the spin-orbit coupling as a perturbation, at first order
the two states corresponding to the Zeeman-split ground state
$|\pm\rangle$ are given as a superposition of few unperturbed
Fock-Darwin states and spin states, $|n,\ell\rangle|s\rangle$, with
$n\in\mathbb{N}$ the principal quantum number, $|\ell|\le n$ the
azimuthal quantum number, and $s=\pm 3/2$, for a detailed see
Ref. \cite{Bulaev-PRL07}. In the case of heavy holes it can be shown
\cite{Bulaev-PRL07} that magnetic-dipole transition ($\Delta n=0$,
$\Delta \ell=0$, and $\Delta s=\pm 1$) are forbidden, while, because of
spin-orbit coupling between states with different orbital momenta and
opposite spin orientations, $|0,0,\pm 3/2\rangle$ and
$|1,\pm 1,\mp 3/2\rangle$, electric-dipole transitions
($\Delta n=\pm 1$, $\Delta \ell=\pm 1$, and $\Delta s=0$) are most likely
to occur. Heavy holes are thus affected by the oscillating electric
field component, but not by the magnetic one. EDSR for heavy hole
appears to be an essential tool for the control of spin dynamics and
for the determination of important parameters, as the effective $g$
factor, effective mass $m$, spin-orbit coupling constants, and spin
relaxation and decoherence time. 

The Hamiltonian for the interaction of HHs with a circularly polarized
electric field, that rotates with frequency $\omega$ in the $XY$-plane,
${\bf E}(t)=E(\sin\omega t,-\cos\omega t,0)$, is given by ${\cal
  H}^{\rm E}=(|e|E/m\omega)(P_x\cos\omega t+P_y\sin\omega t)$. The
coupling between the states $|\pm\rangle$ is given by $\langle +|{\cal
H}^{\rm E}|-\rangle=d_{\rm SO}Ee^{-i\omega t}$, where 
\begin{equation}
d_{\rm SO}=(|e|l/2\omega)
(\beta_1^+\omega_++\beta_1^-\omega_-), 
\end{equation}          
is an effective dipole moment of a heavy hole and it depends on
Dresselhaus spin-orbit coupling constant, perpendicular magnetic field
$B_{\perp}$, lateral quantum dot size, and frequency $\omega$ of the
rf electric field. For details on $\beta_1^{\pm}$ and $l$ see
Ref. \cite{Bulaev-PRL07}.   

The effective master equation for the density matrix $\rho_{nm}$, in
the contest of Bloch-Redfield theory, takes the form of Bloch
equations, with a rf field detuned from $\omega_Z$, $\delta_{\rm
  rf}=\omega_Z-\omega$, Larmor frequency $2d_{\rm SO}E/\hbar$, spin
relaxation time $T_1=1/(W_{+-}+W_{-+})$, $W_{nm}$ being the transition
rate from state $m$ to state $n$, decoherence time $T_2=2T_1$, and
equilibrium value of $\rho_z$ without rf field given by
$\rho_z^T=(W_{+-}-W_{-+}T_1)$. 

The coupling energy between a heavy hole and an oscillating field is
given by
\begin{equation}
\langle {\cal H}^{\rm E}(t)\rangle=-{\bf d}_{\rm SO }\cdot{\bf E}(t),
\end{equation} 
where ${\bf d}_{\rm SO}=d_{\rm SO}(i\rho_{-+}-i\rho_{+-}+\rho_{-+},0)$
is the dipole moment of a heavy hole. The rf power $P=-d\langle{\cal
  H}^{\rm E}(t)\rangle/dt=-\omega d_{\rm SO}E\rho_-$ absorbed by a
heavy hole spin system in a stationary state is given by
\cite{Abragam61}
\begin{equation}
P=\frac{2\omega(d_{\rm SO}E)^2T_2\rho_z^T/\hbar}
{1+\delta_{\rm rf}^2T_2^2+(2d_{\rm SO}E/\hbar)^2T_1T_2}.
\end{equation}
The dependence of $P$ on perpendicular magnetic field $B_{\perp}$ and
frequency of the oscillating field $\omega$ shows three resonances and
one resonant dip. The first resonance corresponds to the Zeeman
energy of the heavy hole $B_{\perp}=\hbar\omega/g_{\perp\mu_B}$, the
second to the first anticrossing between the unperturbed $E_{0,-3/2}$ and
$E_{1,3/2}$ energy levels, $\omega_Z-\omega_-$, the third resonance
reflects the peak in the decoherence rate $T_2^{-1}$ due to an applied
in-plane magnetic field at the second anticrossing
$\omega_Z=2\omega_-$. The resonant dip takes place at zero dipole
moment. 

The study of the position of these resonance allows to
determine $g_{\perp}$, $m$, and $\omega_0$, while the shape and
height provide information about the spin-orbit interaction constants
$\alpha$, $\beta$, and spin-orbit interaction strength due to in-plane
magnetic field. Besides, it is possible to extract informations about
the dependence of spin relaxation and decoherence times on
$B_{\perp}$.

\subsection{Superconducting qubits}

Besides spin qubits in semiconductor quantum dots,
superconducting qubits represent a category of promising candidates
for the implementation of artificial two-level systems as qubits. The 
key ingredient in building superconducting qubits is the strong
nonlinearity of the current-voltage relation of a Josephson
junction. The ability to isolate few charge states on a
superconducting island, together with the possibility to let them
interact through the coherent tunneling of Cooper pairs through the
junction, represent a promising way to control a operate a purely
quantum system (charge qubits). The flux quantization together with
the strong nonlinear potential, arising from the current-voltage
relation, provide a way to isolate few current states and coherently
superimpose them (flux qubit).  

Superconducting qubits can be included in a more general framework of
quantum circuits, that are electrical circuits showing, in the low
temperature regime, quantum behavior, including quantum fluctuations
\cite{Devoret-97}. In this 
context, as $LC$-circuits provide electrical realizations of quantum
harmonic oscillators, showing a linear current-voltage relation,
Josephson junctions provide a full anharmonic counterpart, showing
a more rich spectrum, with groups of few energy levels well separated
from higher bands of the spectrum. 

Several types of superconducting qubits based on Josephson junctions
have been so far theoretically proposed and experimentally
realized (for comprehensive reviews see
\cite{Makhlin-01,Devoret-condmat04}). Apart 
from the particular design of each device, superconducting qubits can
be classified according of the working regime of the Josephson
elements that constitutes the circuit. Every Josephson junction is
characterized by two features: i) a critical current $I_c$, that is
the maximal supercurrent that can flow through the junction; and ii)
an effective capacitance that 
the two superconducting faces have to accumulate charge. Together the
energy associated with the critical current $E_J=I_c\Phi_0/2\pi$ and
the charging 
energy of the associated capacitance $E_C=e^2/2C$ are the two most important
parameters that determine the qubit working regime. For $E_C\gg E_J$
the charge degrees of freedom are well defined and the number of
Cooper pairs in a superconducting island is a well defined quantum
number. Qubits that work in this regime are called charge qubits
\cite{Averin-SolidSC98,Makhlin-Nature99,Nakamura-Nature99,
Pashkin-Nature03,Shnirman-PRL97,Vion-Science02,Wallraff-Nature04,
Blais-PRA04,Schuster-PRL05,Blais-PRA07}. 
To the contrary, for  
$E_C\ll E_J$ flux degrees of freedom have well defined values, and
current states are well defined. Qubits that operate in this regime
are called flux qubits \cite{Orlando-PRB99,Mooij-Science99,
vanderWal-Science00,Chiorescu-Science03}. Other realizations of
superconducting qubits, for different values of the ratio $E_J/E_C$,
and many kind of possible accessible parameter regimes have been
explored. The so called phase qubit
\cite{Ioffe-Nature99,Martinis-PRL02} operates in the flux regime, but
is completely represented by the superconducting phase, and it has no
magnetic flux or circulating current associated. The quantronium
\cite{Cottet-PhysikaC02}, consisting of a split Cooper pair box
arranged in a loop containing an extra large junction for the
read-out.

Experimental observation of Rabi oscillations in driven quantum
circuits have shown several periods of coherent oscillations,
confirming, to some extent, the validity of the two-level
approximation and possibility 
of coherently superimpose the computational two states of the system.
Nevertheless, the unavoidable coupling to a dissipative environment
surrounding the circuit represents a source of relaxation and
decoherence that limit the performances of the qubit for quantum
computation tasks. Therefore, for the implementation of superconducting
circuits as quantum bits, it is necessary to understand the way the
system interacts with the environmental degrees of freedom, and to
reduce their effect, if possible.

\subsection{Circuit theory and system Hamiltonian}

A systematic approach to obtain the Lagrangian of a generic circuit
containing many different lumped elements, as well as Josephson
junctions, has been proposed in
\cite{Burkard-PRB04,Burkard-CircTh-PRB05}. In 
this way, it is possible to construct the full classical Hamiltonian
of the system, quantize it and study its quantized spectrum, in the
two-level approximation.   

\subsubsection{Network graph theory and the equations of motion}

By means of classical network theory, an electric circuit is
represented by an oriented graph, consisting of nodes and
branches. Each branch correspond to a single two-terminal element,
such as resistor, capacitor, inductor, current source, voltage source,
etc.. The branches are then divided into two groups, the tree,
representing a set of branches of the graph connecting all nodes
without containing any loop, and the chords, represented by all the
rest of branches. This way every time a chord branch is added to the
tree a loop is obtained. The grouping in chords and tree depends on
the formalism adopted, that in turn is functional to the kind of
circuit described, either a flux qubit or a charge qubit. All the the
topological information of the circuit is contained in the fundamental
loop matrix
${\bf F}$, that connects tree branches and loops (i.e. chords), such
that the matrix elements $F_{XY}$ can be 1, -1, 0, depending whether
the tree branch $X$ and the chord branch $Y$ have the same orientation,
different orientation in the loop, or do not belong to the same loop. 

The equations of motion are represented by Kirchhoff's laws, and can be
at once written as
\begin{equation}
{\bf F}{\bf I}_{\rm ch}=-{\bf I}_{\rm tr},\qquad
{\bf F}^T{\bf V}_{\rm tr}={\bf V}_{\rm ch}-\dot{\bf \Phi}_{\rm ext}.
\end{equation}
Here
${\bf \Phi}_{\rm ext}$ takes into account the possibility of having
time dependent applied external fluxes. The fluxes and charges of the
circuit represent the canonical variables of system, and they can be
formally defined for the generic element $X$ as
\begin{equation}
{\bf I}_X(t)=\dot{\bf Q}_X(t),\qquad
{\bf V}_X(t)=\dot{\bf \Phi}_X(t).
\end{equation} 
From the last equation and from the second Josephson relation, it is
possible to identify the formal flux associated to the Josephson
junction as the superconducting phase difference
$\boldsymbol{\varphi}$ across the junction, 
\begin{equation}
\frac{{\bf \Phi}_J}{\Phi_0}=\frac{\boldsymbol{\varphi}}{2\pi},\qquad 
{\bf I}_J={\bf I}_c{\bf sin}\boldsymbol{\varphi},
\end{equation}
where the second formula represents the first Josephson relation. 
With current-voltage relations for the various types of other
branches, it is possible to obtain the classical equations of motion
for the superconducting phases  
\begin{equation}\label{Eq:eq-motion-circuit}
{\bf C}\ddot{\boldsymbol{\varphi}}=-{\bf L}_J^{-1}
{\bf \sin}\boldsymbol{\varphi}-{\bf M}_0\boldsymbol{\varphi}
-{\bf M}_d\ast\boldsymbol{\varphi}-\frac{2\pi}{\Phi_0}({\bf N}
{\bf \Phi}_{\rm ext}+{\bf S}{\bf I}_B),
\end{equation}
where ${\bf L}_J^{-1}=2\pi{\bf I}_c/\Phi_0$ is a diagonal matrix for
the Josephson inductances of the junctions, ${\bf M}_0$ is the matrix
of linear inductance, describing their energy and mutual interaction,
${\bf N}$ and ${\bf S}$ describing the inductive coupling of the
phases $\boldsymbol{\varphi}$  with external fluxes and currents,
respectively. ${\bf M}_d(t)$ is a symmetric matrix containing all the
dissipative dynamics of $\boldsymbol{\varphi}$, 
\cite{Burkard-PRB04}.

\subsubsection{Two-level approximation}

Dissipative elements present in the circuit are incompatible with a
Hamiltonian description of the system, therefore for the moment we
omit them. 
In order to derive the Lagrangian for the electric circuit, a complete
set of unconstrained flux and charge degrees of freedom has to be
isolated, such that every assignment of values to those charges and
fluxes represents a possible dynamical state of the system. The
Hamiltonian of the circuit follows straightforwardly from the
Lagrangian by means of a Legendre transformation, and can be formally
written as
\begin{eqnarray}
{\cal H}&=&\frac{1}{2}({\bf Q}-C_{\bf V}{\bf V})^T{\bf C}^{-1}
({\bf Q}-C_{\bf V}{\bf V})+
\left(\frac{\Phi_0}{2\pi}\right)^2U(\boldsymbol{\varphi}),\\
U(\boldsymbol{\varphi})&=&
-\sum_i\frac{2\pi I_{c;i}}{\Phi_0}\cos\varphi_i+
\frac{1}{2}\boldsymbol{\varphi}^T{\bf M}_0\boldsymbol{\varphi}+
\frac{2\pi}{\Phi_0}\boldsymbol{\varphi}^T
({\bf N}{\bf \Phi}_{\rm ext}+{\bf S}{\bf I}_B),
\end{eqnarray} 
where ${\cal C}$ is the capacitance matrix, collecting all the
capacitive elements of the circuit, and describing the effective
charge energy of the system, $C_{\bf V}$ describes the coupling of the
charges ${\bf Q}$ to externally applied voltages ${\bf V}$, 
The number of Cooper pair, that accumulates on a junction capacitance,
and the phase of the superconducting order parameter through the
junction, for sufficiently low temperatures, become quantized and
satisfy canonical commutation rules,
\begin{equation}
\left[\hat{\Phi}_i,\hat{Q}_j\right]=
\left[\frac{\Phi_0}{2\pi}\hat{\varphi}_i,2e\hat{N}_j\right]
=i\hbar\delta_{ij},
\end{equation}
where $2e$ is the charge of a Cooper pair, and $\Phi_0/2\pi$ is flux
quantum. Therefore, once a Hamiltonian is obtained from circuit theory,
its quantization follows straightforwardly.  
The energy of an isolated system is a conserved quantity, therefore
strictly speaking the Hamiltonian should be time independent. However
time-dependent circuit elements, such as alternating currents and
voltages, can be included in the Hamiltonian description as
time-dependent parameters.

Care should be taken when dissipative elements such as resistors are
present in the circuit. In this case a more general approach must be
adopted, in which the system considered is coupled to a environmental
bath, and the dynamics of the circuit under analysis is obtain as the
dynamics of part of a larger isolated system, as discussed in section
\ref{Sec:Spin-boson}.    

Once the Hamiltonian has been obtained and quantized it is possible to
study the temperature regime, in which few low energy states are taken
into consideration. A two-level approximation can be carried out by
considering only the ground state and first excited state, and
neglecting higher levels of the spectrum. The goodness of the
two-level approximation is controlled by the ratio of the temperature
and the energy gap between the first and second excited state,
$k_BT/\Delta_{12}\ll 1$. The Hamiltonian of the two-level system can
be therefore expressed in the form of a pseudo-spin 1/2 
\begin{equation}
{\cal H}=\frac{\Delta}{2}\sigma_x+\frac{\epsilon}{2}\sigma_z,
\end{equation}
where $\Delta$ denotes the tunnel coupling between the two qubit
states $|0\rangle$ and $|1\rangle$, eigenstates of $\sigma_z$, and
$\epsilon$ the bias, due to asymmetry.

\subsection{Decoherence in superconducting qubits}

In this section we choose to describe decoherence effects in only two
realizations of superconducting qubits, namely the phase qubit
\cite{Ioffe-Nature99,Martinis-PRL02} and the flux qubit
\cite{Orlando-PRB99,Burkard-PRB04,Koch-PRB05}. An extensive treatment
of decoherence in the quantronium circuit is carried out in
\cite{Ithier-PRB05}. 

\subsection{The superconducting persistent current qubit: Delft qubit}

\begin{figure}
 \begin{center}
 \includegraphics[width=7cm]{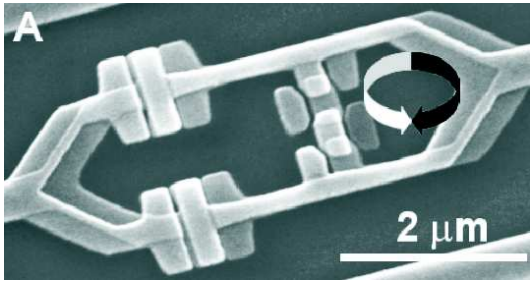}
    \caption{Scanning electron micrograph of the Delft flux qubit
      (small loop with three Josephson junctions) and attached SQUID
      (large loop) \cite{Chiorescu-Science03}
      (With permission from the Authors).
    \label{Fig:FluxQubit}}
 \end{center}
\end{figure}
In the working regime $E_J\gg E_C$, three types of circuit designs
have been proposed, the Delft flux qubit
\cite{Orlando-PRB99,Chiorescu-Science03}, the IBM flux qubit
\cite{Burkard-PRB04}, and its gradiometer variety
\cite{Koch-PRB05}. The phase qubit operates in the same regime, see
Sec.~\ref{Sec:phaseQubit}  

The flux qubit realized at Delft \cite{Chiorescu-Science03} consists
of a superconducting loop interrupted by three Josephson
junctions (see Fig.~\ref{Fig:FluxQubit}). The strong flux regime
$E_J\gg E_C$ allows flux 
quantization of the flux through the loop,
$\varphi_1+\varphi_2+\varphi_3=2\pi n$. Therefore, only two of the
three phases of the junctions play the role of dynamical
variables. For sufficiently low temperatures, in the small loop
inductance limit, the inductive degree of freedom associated with the
loop is frozen, and the effective potential $U(\boldsymbol{\varphi})$
is periodic and shows a double well shape. The charging energy of the
system here plays the role of the kinetic energy, and the Hamiltonian is
written as
\begin{equation}
{\cal H}=-2e^2{\bf \nabla}^T_{\boldsymbol{\varphi}}{\cal C}^{-1}
{\bf \nabla}_{\boldsymbol{\varphi}}+
\left(\frac{\Phi_0}{2\pi}\right)^2U(\boldsymbol{\varphi}).
\end{equation}

The lowest energy states are two flux states localized in the two well
minima $\boldsymbol{\varphi}_L$ 
and $\boldsymbol{\varphi}_R$,  and they correspond to clockwise and 
counter clockwise circulating currents in the loop, $|L\rangle$ and 
$|R\rangle$, encoding the logical $|0\rangle$ and $|1\rangle$ states
of the qubit.   
Tunneling through the potential barrier between the wells lifts the
degeneracy between the two current states, giving rise to a splitting
$\Delta=\langle L|{\cal H}|R\rangle$ between the lowest states
of the system, that become the symmetric and symmetric superpositions
of the current states. An external bias flux can create asymmetry in the
double well, $\epsilon=\langle L|{\cal H}|L\rangle-\langle R|{\cal H}
|R\rangle$. The qubit Hamiltonian written in the
$\{|L\rangle,|R\rangle\}$ basis takes the form 
\begin{equation}\label{Eq:FluxQubitHam}
{\cal H}=\frac{\Delta}{2}\sigma_x+\frac{\epsilon}{2}\sigma_z.
\end{equation}

\subsubsection{Markovian dynamics due to dissipative circuitry}

The regime of working of flux qubits, in which the charging energy is
much smaller than the Josephson energy, $E_C\ll E_J$, makes the flux
qubits substantially insensitive to background charge fluctuations.  
Still, however, other mechanisms can affect their phase coherence and
in order to implement them as building blocks for quantum computation
schemes, it is necessary to understand which sources of decoherence
affect the short time dynamics of flux qubits and reduce as much as
possible their effect. 

Several sources of dissipation for flux qubits have been discussed
throughout the literature \cite{Tian-00}, background charge fluctuations
($\tau_{\varphi}\approx 0.1 {\rm s}$), as well as quasiparticle
tunneling in the superconductor with a non-vanishing subgap conductance 
($\tau_{\varphi}\approx 1 {\rm ms}$). Nuclear spins in the substrate
have also been considered as a possible source of dissipation. Static
random magnetic field produced by the nuclear spins may induce shifts
in qubit frequencies, but no dephasing is expected until a typical
nuclear relaxation time, which can be very long, up to minutes, due to
the slow dynamics of nuclear spins. 

However, the most efficient source of dissipation for flux qubits is
represented by fluctuations in the external circuit that produce
fluctuating magnetic fluxes through the loop. The coupling of flux
degrees of freedom of the qubit to the dissipative environmental
elements is well described in the graph formalism described in
\cite{Burkard-PRB04}. In the Born-Markov approximation, it can be
shown that the Redfield tensor, written in the eigenbasis
$\{|n\rangle\}$ of ${\cal H}_S$, is entirely determined by  
\begin{eqnarray}
{\rm Re}\Gamma^{(+)}_{lmnk}&=&({\bf m}\cdot\boldsymbol{\varphi})_{lm}
({\bf m}\cdot\boldsymbol{\varphi})_{nk}J(|\omega_{nk}|)
\frac{e^{-\beta\omega_{nk}/2}}{\sinh\beta|\omega_{nk}|/2},
\label{Eq:ReGamma}\\
{\rm Im}\Gamma^{(+)}_{lmnk}&=&-({\bf m}\cdot\boldsymbol{\varphi})_{lm}
({\bf m}\cdot\boldsymbol{\varphi})_{nk}J(|\omega_{nk}|)
\frac{2}{\pi}P\int_0^{\infty}d\omega\frac{J(\omega)}{\omega^2-\omega_{nk}^2}\\
&\times&\left(\omega-\omega_{nk}\coth\frac{\beta\omega}{2}\right),
\label{Eq:ImGamma}
\end{eqnarray}
where $\beta=k_BT$ and the ${\bf m}\cdot\boldsymbol{\varphi}$ appears
in the Hamiltonian for the system-bath coupling, ${\bf m}$ being
related to the topology of the dissipative circuitry. 

In the two-level approximation, the rates for transitions from the
two-level subspace to higher states can be neglected, and Eqs.
(\ref{Eq:ReGamma}) and (\ref{Eq:ImGamma}) simplify and the dynamics of
the $2\times 2$ density matrix of the system can be cast in the form
of Bloch equations for the dynamics of a pseudo spin 1/2. In this
framework the relaxation matrix contains just two rates, $T_1^{-1}$
and $T_2^{-1}$ for the decay of the longitudinal and transverse pseudo
spin component, respectively. The latter in turn is limited by
relaxation time $T_1$ and pure dephasing time $T_{\phi}$,
$1/T_2=1/2T_1+1/T_{\phi}$, and the two rates are given by
\cite{Burkard-PRB04}  
\begin{eqnarray}
\frac{1}{T_1}&=&4|\langle 0|{\bf
  m}\cdot\boldsymbol{\varphi}|1\rangle|^2
J(\omega_{01})\coth\frac{\omega_{01}}{2k_BT},\\
\frac{1}{T_{\phi}}&=&|\langle 0|{\bf
  m}\cdot\boldsymbol{\varphi}|0\rangle-\langle 1|{\bf
  m}\cdot\boldsymbol{\varphi}|1\rangle|^2\left.
\frac{J(\omega)}{\omega}\right|_{\omega\rightarrow 0}2k_BT.
\end{eqnarray}  
Typically, $T_{\phi}$ can be made to diverge for an appropriate choice 
of external fluxes such that $\langle 0|{\bf m}\cdot
\boldsymbol{\varphi}|0\rangle=\langle 1|{\bf m}\cdot
\boldsymbol{\varphi} |1\rangle$. However, this divergence is not
expected to show up experimentally, since it will be cut off by other
mechanisms. 

The two lowest energy states, eigenstates of the Hamiltonian
Eq.~(\ref{Eq:FluxQubitHam}), are given by 
\begin{eqnarray}
|0\rangle&=&\frac{1}{\sqrt{2}}\left(\sqrt{1+
\frac{\epsilon}{\omega_{01}}}|L\rangle+\sqrt{1-
\frac{\epsilon}{\omega_{01}}}|R\rangle\right),\\
|1\rangle&=&\frac{1}{\sqrt{2}}\left(\sqrt{1-
\frac{\epsilon}{\omega_{01}}}|L\rangle-\sqrt{1+
\frac{\epsilon}{\omega_{01}}}|R\rangle\right),
\end{eqnarray} 
where $\omega_{01}=\sqrt{\epsilon^2+\Delta^2}$. Approximating
the localized flux states $|L\rangle$ and $|R\rangle$ as Gaussians
centered in the minima of the double well, the relaxation rate
$T_1^{-1}$ and the pure dephasing rate $T_{\phi}^{-1}$ are given by
\begin{eqnarray}
\frac{1}{T_1}&\approx&\left(\frac{\Delta}{\omega_{01}}\right)^2
|{\bf m}\cdot\Delta\boldsymbol{\varphi}|^2
\left(1+\frac{S^2}{2}\right)^2J(\omega_{01})
\coth\frac{\omega_{01}}{2k_BT},\\
\frac{1}{T_{\phi}}&\approx&\left(\frac{\epsilon}{\omega_{01}}\right)^2
|{\bf m}\cdot\Delta\boldsymbol{\varphi}|^2
\left(1+\frac{S^2}{2}\right)^2
\left.\frac{J(\omega)}{\omega}\right|_{\omega\rightarrow 0}2k_BT,
\end{eqnarray} 
where $S=\langle L|R\rangle$ is the overlap between the two
Gaussians. The vector $\Delta\boldsymbol{\varphi}$ connects the two
minima of the double well. These relation are valid in the Markov
limit and at in Born approximation, where the system-bath interaction
is considered only at first order. By inspection of the previous
formula it is clear that a symmetric double well potential, for which
$\epsilon=0$, let the dephasing time to diverge. This is realized for
a value of the external applied magnetic flux $\Phi_{\rm
  ext}=\Phi_0/2$, being $\epsilon\propto(\Phi_{\rm
  ext}/\Phi_0-1/2)$. Moreover for ${\bf
  m}\cdot\Delta\boldsymbol{\varphi}=0$ the environment is decoupled
from the system, and both the relaxation and dephasing time diverge. 

In Ref.~\cite{Burkard-PRB04} an estimate of the leakage rate due to
transition from the qubit states $k=0,1$ to higher energy levels
$n=,2,3,.\ldots$ outside the qubit subspace can be quantified from
Eqs.(\ref{Eq:ReGamma}) and  (\ref{Eq:ImGamma}),
\begin{equation}
\frac{1}{T_{L,k}}=4\sum_n|\langle n|{\bf m}\cdot\boldsymbol{\varphi}
|k\rangle|^2J(\omega_{kn})\coth\frac{\omega_{kn}}{2k_BT}.
\end{equation} 
In the regime $\eta\gg \Delta,\delta,\epsilon$, where $\eta$ is the
energy splitting between the lowest two states $|L\rangle$ and
$|R\rangle$ and the third energy level, and $\delta$ is the coupling
between the qubit subspace and the next higher level, the dominant
leakage occurs with rate 
\begin{equation}
\frac{1}{T_L}\approx 4\left(\frac{\delta}{\eta}\right)^2
|{\bf m}\cdot\Delta\boldsymbol{\varphi}|^2
J(\eta)\coth\frac{\eta}{2k_BT},
\end{equation}
and in the regime in which the two-level approximation is well
defined, $\Delta\gg k_BT$, thermally activated leakage is strongly
suppressed.

\subsubsection{Thermal photon noise induced dephasing}

Besides magnetic flux fluctuations, an important source of dephasing
is represented by thermal photon noise in the read-out part of the
circuit. To measure the state of the qubit a superconducting quantum
interference device (SQUID) is coupled to the qubit via mutual
inductance. When the SQUID is biased by a current that has a value
smaller than the critical current, the SQUID acts just as an effective
inductor, whose linear inductance depends on the qubit current
state. The critical current at which the SQUID switch to a normal
state can have two values $I_c^{|0\rangle}$ and $I_c^{|1\rangle}$,
according to the two qubit possible current states. A bias current
pulse of amplitude $I_B$, $I_c^{|0\rangle}<I_B<I_c^{|1\rangle}$,
allows to discriminate the two qubit states. The read-out apparatus
consisting of a dc-SQUID and a shunt capacitor $C_s$ form a weakly
damped harmonic oscillator of frequency $\omega_{\rm ho}$, that is
detuned from the qubit frequency 
$\omega_q$. The presence of $n$ photons in the harmonic oscillator
induces a shift in the qubit frequency,
$\omega_{q,n}-\omega_{q,0}=n\delta\omega_0$, where the shift per
photon depends on the effective qubit-oscillator coupling. Assuming
that the pure dephasing time $\tau_{\phi}$ is much larger that the
inverse of the damping rate $\kappa$, $\tau_{\phi}\gg 1/\kappa$, 
thermally excited photons in the oscillator produce a dephasing 
\cite{Blais-PRA04,Bertet-condmat05,Bertet-PRL05}
\begin{equation}\label{Eq:PhotonIndDephTime}
\tau_{\phi}=\frac{\kappa}{\bar{n}(\bar{n}+1)\delta\omega_0^2}, 
\end{equation}    
where $\bar{n}=(\exp(\hbar\omega_{\rm ho}/k_BT)-1)^{-1}$ is the
thermal average number of photon in the oscillator. A similar effect
has been observed in an experimental work \cite{Schuster-PRL05} in
which a charge qubit is coupled to a superconducting waveguide
resonator, slightly detuned from the qubit frequency. There, opposite
to the case here described, the oscillator is driven and a shift and a
broadening in the qubit resonance frequency appears, as a consequence
of an ac-Stark shift and of photon shot noise.  

The flux qubit of Ref.~\cite{Bertet-PRL05} has been engineered with
four Josephson junction to ensure a symmetric qubit-SQUID coupling
\cite{Burkard-PRB05}. In the usual design of the Delft qubit
\cite{Chiorescu-Science03}, the two symmetric arms of the SQUID render
the qubit immune to bias current $I_B$ fluctuations. At zero dc bias,
$I_B=0$, a small fluctuating current caused by the finite impedance of
the external controls is divided equally into two branches of the
SQUID loop and no net current flows through the three-Josephson
junctions of the qubit line. However, the double layer structure of
the Josephson junctions,  being an artefact of the shadow evaporation
technique used to construct Josephson junctions, induces asymmetry in
the circuit. Using a forth much larger Josephson junction, for which
the Josephson energy can be usually neglected, symmetry in
the double layer structure is restore and effects of fluctuations in
the bias current are suppressed.       

The qubit energy bias can be written as the sum of two contributions,
$\epsilon=\eta+\lambda$, where $\eta=2I_p(\Phi_{\rm ext}-\Phi_0/2)$
($I_p$ is the qubit persistent current) is controlled by the external
flux $\Phi_{\rm ext}$ and $\lambda=2I_pMJ(I_B)/h$ which depends on
$I_B$ via the SQUID circulating current. This dependence has two
crucial consequences: first the qubit bias point $\Phi_{\rm ext}^*$
for which $\partial \omega_q/\partial \Phi_{\rm ext}=0$ results
shifted by the measurement pulse. Therefore it is possible to operate
the qubit at 
the flux-insensitive point, while keeping a difference in the
expectation value of the current in the two qubit states, which is
a crucial requirement for measuring the qubit state. Second a coupling
between the qubit and the external harmonic oscillator, the so-called
plasmon mode, arise, with an interaction Hamiltonian 
\begin{equation}
{\cal H}_{\rm q-ho}\propto[g_1(I_B)(a+a^{\dag})
+g_2(I_B)(a+a^{\dag})^2]\sigma_z,
\end{equation}
where $g_1(I_B)\propto(d\lambda/dI_B)$ and
$g_2(I_B)\propto(d^2\lambda/dI_B^2)$ \cite{Bertet-condmat05}. 
For a particular $I_B^*$ that realizes $d\lambda/dI_B=0$, it is
possible to switch $g_1(I_B)$ off \cite{Bertet-PRL05,Burkard-PRB05}. 

Working at these optimal point, the qubit is immune from external flux
and bias current fluctuations at first order, $\partial
\omega_q/\partial\Phi_{\rm ext}(\Phi_{\rm ext}^*,I^*_B)=
\partial\omega_q/\partial I_B(\Phi_{\rm ext}^*,I^*_B) =0$. The shift
per photon of $\delta\omega_0$ is given, at second order perturbation
theory in ${\cal H}_{\rm q-ho}$ \cite{Bertet-condmat05,Bertet-PRL05},
by 
\begin{equation}
\delta\omega_0=4\left[\left(g_1(I_B)\sin\theta\right)^2
\frac{\omega_q}{\omega^2_{\rm ho}-\omega_q^2}-
g_2(I_B)\cos\theta\right], 
\end{equation}
where $\cos\theta=\epsilon/\omega_q$. For some value
$\epsilon^*(I_B)<0$ one obtains $\delta\omega_0=0$. In
Ref.~\cite{Bertet-PRL05}, via spectroscopy the authors demonstrated the
existence of a line $\epsilon^*(I_B)$, that includes $I_B=I_B^*$ and
$\epsilon=0$, providing an optimal point with respect to bias current
noise, flux noise, and photon noise. Measurements of the qubit
spectral line shape at the optimal point showed, for the particular
sample Ref.~\cite{Bertet-PRL05}, a twin peak structure, which could
arise from strong coupling to one microscopic fluctuator. An effective
dephasing time $t_2=2/\pi(w_1+w_2)$ is obtained by fitting the peaks
with two Lorentzians of width $w_1$ and $w_2$. Measurements of the
spin-echo decay time $T_{\rm echo}$, particularly indicated in case of
relatively high frequency noise, as photon noise in the plasma mode
that occurs at $\kappa\approx 130{\rm MHz}$, gives at the optimal
point $T_{\rm echo}=3.9 \mu{\rm s}$. By studying the variation of
$T_{\rm echo}$ and $t_2$ as a function of $\epsilon$, a sharp peak is
found at $\epsilon=0$ for $I_B=I_B^*$, while for $I_B=0$ the peak
shifts towards $\epsilon<0$. The variation of the maximum in $t_2$ as
a function of $I_B$ show that the maximal coherence time is not
obtained at $\epsilon=0$, as it would be expected for flux or bias
current noise. On the other hand, it fits with the line
$\epsilon^*(I_B)$ for which $\delta\omega_0=0$, suggesting that
thermally induced photon noise, rather than flux noise or bias current
noise, is responsible for the qubit dephasing. For a temperature
$T=70{\rm mK}$ and quality factor $Q=150$, which yields a mean photon
number $\bar{n}=0.15$, the dephasing time $\tau_{\phi}$
Eq.~(\ref{Eq:PhotonIndDephTime}) closely matches the spin-echo
measurements. 

\subsection{The superconducting phase qubit}
\label{Sec:phaseQubit}

\begin{figure}
 \begin{center}
 \includegraphics[width=9cm]{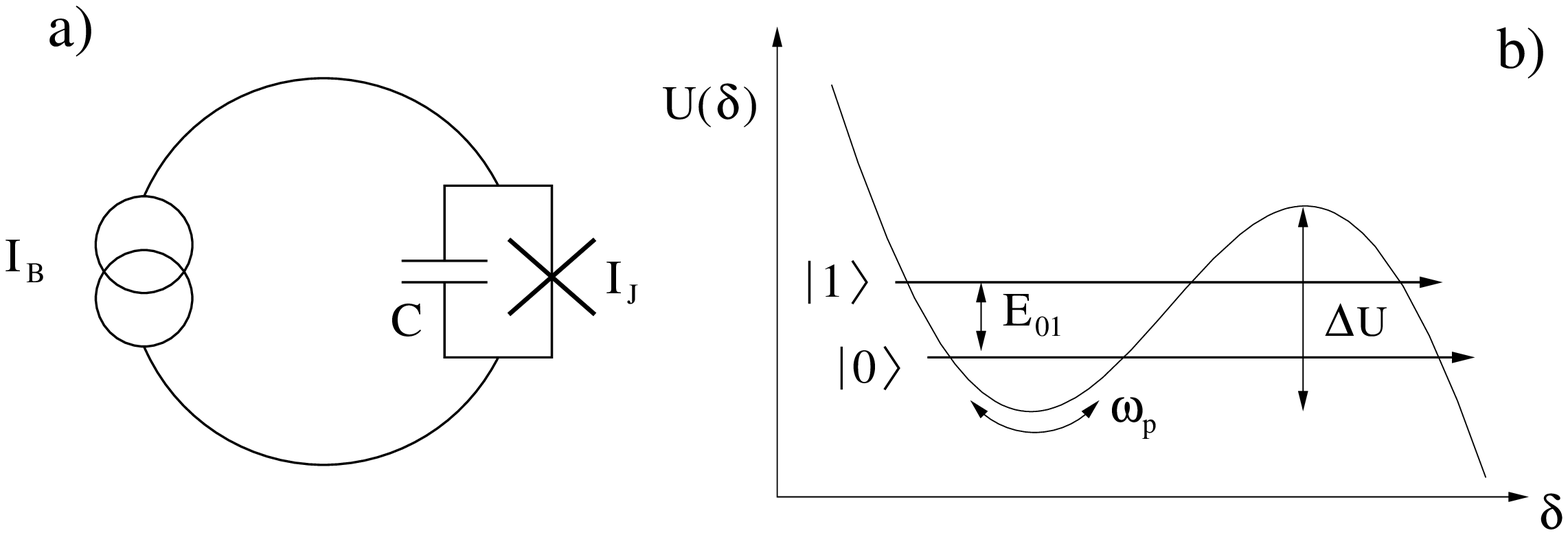}
    \caption{a) Schematic representation of the phase qubit circuit,
      constituted by a current-biased Josephson junction. b)
      Anharmonic potential $U(\delta)$ showing the two lowest energy
      states $|0\rangle$ and $|1\rangle$, separated by an energy
      splitting $E_{01}$. The plasma frequency $\omega_p$ is given by
      the local quadratic curvature of the potential at the bottom of
      the well, and $\Delta U$ is the potential barrier that separates
      the two energy levels in the well from a continuum of levels on
      the right side of the barrier. \label{Fig9}}
 \end{center}
\end{figure}
The phase qubit works in a regime in which $E_J\gg E_C$ and the circuit
consists of a loop with a single large Josephson junction, as shown in
Fig.~\ref{Fig9}~(a). The circuit is biased with a current $I$
typically driven close to the critical current $I_0$ of the
junction. The Hamiltonian of the system is
\begin{equation}
H=\frac{\hat{Q}^2}{2C}-\frac{I_0\Phi_0}{2\pi}\cos\hat{\delta}
-\frac{I\Phi_0}{2\pi}\hat{\delta}
\end{equation}
where $\Phi_0/2\pi=\hbar/2e$ is the superconducting flux
quantum. Charge and phase operators, $\hat{Q}$ and $\hat{\delta}$,
that correspond to the charge on the Josephson junction capacitance and
the superconducting phase across the Josephson junction respectively,
are conjugate variables that satisfy the canonical commutation rule
$[\hat{\delta},\hat{Q}]=2ei$. For a large area junction
$I_0\Phi_0/2\pi=E_J\gg E_C=e^2/2C$ the superconducting phase has a
well defined value and quantum mechanical behavior can be observed.   
The Josephson inductance and the junction capacitance form
an anharmonic ``LC'' resonator. The anharmonic potential as a function
of the superconducting phase across the junction can be approximated
by a cubic potential parametrized by the potential barrier $\Delta
U(I)=(2\sqrt{2I_0\Phi_0/3\pi})[1-I/I_0]^{3/2}$  and a classical plasma
oscillation frequency at the bottom of the potential well
$\omega_p(I)=2^{1/4}(2\pi I_0/\Phi_0C)^{1/2}[1-I/I_0]^{1/4}$. The two
qubit states $|0\rangle$ and $|1\rangle$ are encoded in the two lowest
quantum states in the potential well, and have energy splitting
$E_{01}=\hbar\omega_p(1-5\hbar\omega_p/36\Delta U)$. Unlike in a flux
or charge qubit, in a phase qubit the state ($|0\rangle$ or
$|1\rangle$) is exclusively distinguished by the phase wavefunction,
and not by any macroscopic quantity, such as current or charge. 
Transitions are driven by applying microwaves at frequency 
$\omega_{01}=E_{01}/\hbar$. For more details on the phase qubit we
refer to Ref.~\cite{Martinis-PRL02}. 

Coherent control of the qubit is obtained through the bias current 
\begin{equation}
I(t)=I_{\rm dc}+I_{1/f}(t)+I_{\mu wc}(t)\cos(\omega_{01}t)+
I_{\mu ws}(t)\sin(\omega_{01}t),
\end{equation}
where $I_{1/f}$, $I_{\mu wc}$, and $I_{\mu ws}$ are varied in
time slowly compared to $2\pi/\omega_{01}$. In the frame rotating with
frequency $\omega_{01}$, the qubit Hamiltonian is given by
\begin{equation}
H=\epsilon\frac{\Phi_0}{2\pi}I_{\mu wc}(t)\sigma_x+
\epsilon\frac{\Phi_0}{2\pi}I_{\mu ws}(t)\sigma_y+
\frac{1}{2}\frac{\partial E_{01}}{\partial I_{\rm dc}}I_{1/f}(t)\sigma_z,
\end{equation} 
where $\epsilon=\sqrt{E_C/E_{01}}$, and $\sigma_{x,y,z}$
are the Pauli operators. 

\subsubsection{Decoherence of a phase qubit due to an arbitrary noise
  source}

Since the qubit is controlled by the bias current, noise in the bias
current can represent a source of decoherence for the qubit. In
Ref.~\cite{Martinis-PRB03} a physical picture of decoherence is
presented for a phase qubit.  
 
In a Bloch picture, the state of the qubit is represented by
$|\psi\rangle=\cos(\theta/2)|0\rangle+e^{i\phi}\sin(\theta/2)|1\rangle$. 
Low frequency noise induces fluctuations is the longitudinal $z$
component of the pseudo-spin representing the qubit that lead to
dephasing of the qubit. The phase noise after time $t$ is
\begin{equation}
\phi_n(t)=\frac{\partial\omega_{01}}{\partial I_{\rm dc}}
\int_0^tdt'I_n(t'),
\end{equation} 
and it arises from current noise $I_n(t)$. The magnitude of the phase
noise is described by $\langle\phi_n^2(t)\rangle$, and it can be
obtained through the noise spectral density $S_I(f)$,
\begin{equation}
\langle\phi_n^2(t)\rangle=\left(
\frac{\partial\omega_{01}}{\partial I_{\rm dc}}\right)^2
\int_0^{\omega_{01}/2\pi}dfS_I(f)W_0(f),
\end{equation}
where $S_I(f)$ is given by the Fourier transform of noise correlator  
$\langle I_n(t)I_n(0)\rangle=\int_0^{\infty}dfS_I(f)\cos(2\pi ft)$,
the spectral weight $W_0(f)=\sin^2(\pi ft)/(\pi f)^2$, and the
integral on the frequency has been cutoff for frequency greater than
$\omega_{01}/2\pi$. This last assumption is justified since for those
frequencies the noise current flows mainly through the junction
capacitance, rather than the junction itself, thus not substantially
affecting $\omega_{01}$. The magnitude of noise is defined as the
mean-square amplitude of the current noise at frequency $f$ per 1 Hz
bandwidth. For low frequency $f\le 1/t$, $W_0(f)$ is rather constant,
whereas it decreases as $1/f^2$ at higher frequencies. As a
consequence, phase noise affects the qubit dynamics only at low
frequencies for most noise sources. For constant (white) noise
$S^0_I$, one has
\begin{equation}\label{Eq:phi2n}
\langle\phi_n^2(t)\rangle=\left(
\frac{\partial\omega_{01}}{\partial I_{\rm
    dc}}\right)^2\frac{S^0_It}{2}. 
\end{equation}    

At higher frequencies close to $\omega_{01}$, noise induces
transitions between the two qubit states $|0\rangle$ and
$|1\rangle$. The current that controls these transitions is given by
$I_{\mu wc}(t)\cos(\omega_{01}t)+I_{\mu ws}(t)\sin(\omega_{01}t)$, and
mixing from noise around frequency $\omega_{01}$ can be understood as
low frequency noise in $I_{\mu wc}(t)$ and $I_{\mu ws}(t)$. Random
fluctuations along the transverse $x$ and $y$ components of the qubit
induce transitions between the qubit states. For constant spectral
density around $\omega_{01}$ given by $2S_I(\omega_{01}/2\pi)$, an
application of the previous results gives
\begin{equation}
\langle\theta_x^2(t)\rangle=
\langle\theta_y^2(t)\rangle=\frac{E_C}{E_{01}}\left(
\frac{\Phi_0}{2\pi}\right)^2S_I(\omega_{01}/2\pi) t.
\end{equation} 
The random angles $\chi=\phi,\theta_x$ and $\theta_y$ are assumed to be
Gaussian distributed, with zero mean and mean squared noise
$\langle\chi^2\rangle$ previously calculated,
$dp(\chi)/d\chi=\exp(-\chi^2/2\langle\chi^2\rangle)/
\sqrt{2\pi\langle\chi^2\rangle}$.   

If the qubit is initially in the
ground state, that is when it is parallel to the $z$ direction in the
Bloch sphere, when the noise is small, it is immune to phase noise at
low frequency. However, transverse noise around frequency
$\omega_{01}$ can induce transitions in between the qubit states. The
probability $p_0$ to be in the state $|0\rangle$ is given in the Bloch 
picture by
$\cos^2(\theta/2)\simeq\cos^2[\sqrt{\theta_x^2+\theta_y^2}/2]$.
With the values previously obtained for the mean-square noise,
averaging over the Gaussian distribution gives 
\begin{equation}
p_0=\frac{1}{2}\left(1+e^{-t/T_1}\right),\qquad
\frac{1}{T_1}=\frac{E_C}{E_{01}}\left(
\frac{\Phi_0}{2\pi}\right)^2S_I(\omega_{01}/2\pi).
\end{equation}  
The rate $1/T_1$ describes absorption and emission rate for the
stimulated transitions $0\rightarrow 1$ and $1\rightarrow 0$. Since low
frequency noise cannot add energy $\hbar\omega_{01}$, there is no
contribution from phase noise. 

Effects of noise on a superposition state can be understood within a 
``Ramsey fringe'' picture. Through a $\pi/2$ pulse the qubit is
rotated from the ground state $|0\rangle$ to the state
$(|0\rangle+|1\rangle)/\sqrt{2}$, that points in the $x$ direction in
the Bloch sphere, and left evolving for a time $t$,
after which a subsequent $\pi/2$ pulse is performed and the qubit
state is finally measured. During the evolution between the two
$\pi/2$ pulses, the state of the qubit can change due to noise in
$\phi$ and $\theta_y$, therefore, both phase and stimulated
transitions noise affect the qubit dynamics. In this case the total
decoherence rate is given by the Korringa relation \cite{Abragam-61}
$1/T_2=1/T_{\phi}+1/2T_1$, where $1/T_{\phi}$ is directly
extracted from Eq.~(\ref{Eq:phi2n})
$1/T_{\phi}=(\partial\omega_{01}/\partial I_c)^2S_I^0/4$.

\subsection{Decoherence due to $1/f$ noise }

Much effort has been spent recently to understand how noise at low
frequencies affects the dynamics of superconducting qubit, both from a
theoretical and an experimental point of view. In particular,
signatures of non-Markovian dynamics are believed to be due to $1/f$
noise. Both charge and flux $1/f$ noise contribute to decoherence, the
former affecting mostly the dynamics of superconducting qubits based
on the charge degrees of freedom, and the latter affecting flux and
phase qubits. Here we concentrate only on $1/f$ flux noise, and refer
to the literature for the case of $1/f$ charge noise
\cite{Astafiev-PRL06,Yu-arxiv07}.

\subsubsection{Model for $1/f$ flux noise in SQUIDs}

The origin of $1/f$ flux noise in low-$T_c$ devices has not yet been
completely understood. 
Critical-current fluctuations in Josephson junctions are believed to
arise from trapping and release of electrons in traps in the tunnel
barriers. 
In Ref.~\cite{Koch-PRL07}, a model for $1/f$ noise in low-$T_c$
devices is proposed, based on the assumptions that it arises from
thermally activated hopping of unpaired electrons on and off
defects. The spin of the electron is locked in direction while the
electron is trapped, and the directions are randomly change from trap
to trap. Uncorrelated changes of these spin directions give rise to 
random telegraph signals that produce a $1/f$ power spectrum. The
electron is assumed to occupy the low-energy spin direction during the
entire time it resides on the defect. For zero magnetic field ${\bf
  B}$, transitions between the two degenerate Kramers' doublet are
strongly suppressed, implying that direct phonon scattering is
forbidden. On the other hand, the magnetic field is not strictly zero,
and fluctuating dipole fields can arise from neighboring defects.   

The magnetic moment of a defect $\hat{\bf M}=\mu_B(\hat{\bf
  L}+2\hat{\bf S})$ can be locked as a consequence of spin-orbit
coupling, that makes it stable with respect to these weak
fluctuations. The locking effect can be modeled by the Hamiltonian 
\begin{equation}\label{Eq:HamDefect}
H=\sum_{i=x,y,z}V_i|p_i\rangle\langle p_i|+
\lambda\hat{\bf L}\cdot\hat{\bf S}+\mu_B{\bf B}
\cdot(\hat{\bf L}+2\hat{\bf S}).
\end{equation}
The unpaired electron occupies a $p$ orbital, and $V_{x,y,z}$ are the
matrix elements of the crystal-field potential. The spin-orbit
coupling constant, depending on the different kind of defects, can
vary in magnitude within a large range of values, but for defects
involving atomic weights  near that of Si, $|\lambda|\approx 300~{\rm
  K}$. The crystal-field $V_{x,y,z}$ can be at most $\approx 2000~{\rm
K}$. In defects for which $\lambda<0$, ${\bf L}$ and ${\bf S}$ are
parallel and ${\bf M}$ is large, while for defects for which
$\lambda>0$  ${\bf L}$ and ${\bf S}$ are antiparallel, and ${\bf M}$
is close to zero. Therefore, the $\lambda<0$ defects are expected to
be most important for flux noise.  

A random distribution of defects over the substrate is assumed, and
the flux noise coupled into a SQUID by a spatially random distribution
of electron spins, whose orientations fluctuate, is calculated. The
magnetic moment is represented by a small current loop that couples to
the SQUID loop by a mutual inductance $M(x,y)$. The SQUID loop is
schematized like a square frame of inner and outer dimensions of $2d$
and $2D$, and thickness $W=D-d$, lying in the plane $z=1~\mu{\rm
  m}$. The current loop can lie in the $z=0~\mu{\rm m}$ plane,
resulting in a perpendicular moment, or in the $x$ or $y$ plane
centered at $z=0$, resulting in a in-plane moment. The small 
loop current has an effective area $A=(0.1~\mu{\rm m})^2$, and a
current $i$ flows in it, such that $Ai=\mu_B$, with flux per Bohr
magneton given by $\Phi_s/\mu_B=M(x,y)/A$. Perpendicular and in-plane
flux per Bohr magneton show a qualitative opposite behavior as
function of the position in the SQUID loop plane, the former peaking
on the edges of the superconductor, and the latter peaking at the
midpoint of the superconductor, both falling off away from these
points. The total mean square normalized flux noise coupled to the
SQUID is given by 
\begin{equation}
\langle(\delta\Phi_s)^2\rangle=8n\mu_B^2\int_0^{L+D}dx\int_0^xdy
\frac{M^2(x,y)}{A^2},
\end{equation}  
where $n$ is the areal density of defects, and the integral over the
plane is cutoff a distance $L$ away from the SQUID loop. The mean
square noise is given by $\langle(\delta\Phi_{\rm st})^2\rangle=[
\langle(\delta\Phi_{{\rm si},x})^2\rangle+\langle(\delta
\Phi_{{\rm si},y})^2\rangle+\langle(\delta\Phi_{\rm sp})^2\rangle]/3$,
and the spectral density $S_{\Phi}(f)=\alpha/f$ is extracted by 
$\langle(\delta\Phi_{\rm st})^2\rangle=\alpha\int_{f_1}^{f_2}df/f
=\alpha\ln(f_2/f_1)$, and, for $f_2/f_1\approx 10^{13}$, 
\begin{equation}
S_{\Phi}(f)/\Phi_0^2\approx\langle(\delta\Phi_s/\Phi_0)^2\rangle/30f.
\end{equation} 

Noise levels in agreement with the observations are obtained for
$n\approx 10^{17}~{\rm m}^{-2}$. However, they strongly vary with the
geometry of the the SQUID and the tunnel barriers, and with the
fabrication details.

\subsubsection{Decoherence of flux qubits due to $1/f$ noise}

Recently experiments \cite{Yoshihara-PRL06,Kakuyanagi-PRL07} have
reported the behavior of the echo signal in flux qubits at
various bias conditions. The energy splitting depends on the applied
external magnetic flux, and thus it is sensitive to flux noise. As a
result they found that the at the optimal point, where the energy
splitting is insensitive at first order to magnetic flux fluctuations,
the coherence time is limited by energy relaxation processes,
$T_1$, and the dephasing of the flux qubit is mostly determined by the
high-frequency noise $S_{\Phi}(\omega\approx\Delta/\hbar)$. 

In a Markovian approximation the energy relaxation contributes to the
dephasing process via $1/T^{\rm echo}_2=1/2T_1+\Gamma_{\varphi}$,
where $\Gamma_{\varphi}$ is the pure dephasing rate. For dephasing
dominated by magnetic flux noise with a smooth spectrum near
$\omega=0$, the pure dephasing rate is given by
$\Gamma_{\varphi}^{\Phi}\approx(\partial\Delta E/\partial\Phi_{\rm
  ext})^2S(\omega=0)$, where $\Delta E=\sqrt{\epsilon(\Phi_{\rm
    ext})^2+\Delta^2}$. Taking into account the fact that, from their
experimental data, the relaxation time $T_1$ is almost independent on
the external applied magnetic flux, and that close to the optimal
point the $\partial\Delta E/\partial\Phi_{\rm ext}\propto\Phi_{\rm
  ext}$, a parabolic behavior of $1/T^{\rm echo}_2$ is expected in
this region. On the other hand, away from the optimal point, they
observed a linear increasing of $1/T_2^{\rm echo}$ with respect of the
applied external magnetic flux. Therefore, the experimental data
cannot be explained within the framework of Bloch-Redfield decoherence
theory, in which the assumption of short time correlated noise (white
noise around $\omega=0$) holds.  

The experimental observations can be explained with the presence of
$1/f$ flux noise. In Refs.~\cite{Yoshihara-PRL06,Kakuyanagi-PRL07}, in
order to separate the contribution to dephasing due to direct
transitions between energy levels, the echo signal is expressed as
$\rho(t)=e^{-t/2T_1}\rho_{\rm echo}(t)$, with
$\rho(t)=\langle\sigma_z\rangle$. In this case the expectation value
of $\sigma_z$ is given by a non-exponential decay curve, and at the
end of the echo sequence  
\begin{equation}
\rho_{\rm echo}(t)=\left\langle\exp\left\{-iv_{\Phi}
\left[\int_0^{t/2}\Phi(\tau)d\tau-
\int_{t/2}^{t}\Phi(\tau)d\tau\right]\right\}\right\rangle,
\end{equation}
where $v_{\Phi}=(\Phi_0/\hbar)\partial\Delta E/\partial\Phi_{\rm
  ext}$. 
Assuming Gaussian statistics of the fluctuations
of the external flux, the decoherence rate can be expressed through
the noise spectral function,
$S_{\Phi}(\omega)=(1/\pi)\int_0^{\infty}dt\cos\omega
t\langle\Phi(t)\Phi(0)\rangle$,
\begin{equation}\label{Eq:Gaussiandecay1overf}
\rho(t)=\exp\left[-\frac{t^2v_{\Phi}^2}{2}
\int_{-\infty}^{\infty}S_{\Phi}(\omega)
\frac{\sin^4(\omega t/4)}{(\omega t/4)^2}\right]
\end{equation}    
For a $1/f$ spectrum $S_{\Phi}=A_{\Phi}/\omega$ one obtains
\begin{equation}
\rho_{\rm echo}(t)=\exp\left[-(\Gamma_{\varphi}^{\Phi}t)^2\right],
\qquad 
\Gamma_{\varphi}^{\Phi}\equiv|v_{\Phi}|\sqrt{A_{\Phi}\ln 2}.
\end{equation}

In Ref.~\cite{Galperin-PRB07}, motivated by the experimental results
of Refs.~\cite{Yoshihara-PRL06,Kakuyanagi-PRL07}, a theoretical
analysis of the dephasing due to $1/f$ noise in flux qubits has been
presented. The problem is described by choosing spin-fluctuators as
source of low-frequency noise. Random switching of a fluctuator
between its two metastable states gives rise to a random telegraph
noise. Transitions in fluctuators with energy splitting larger than
the temperature are strongly suppressed and only fluctuators
whose energy splitting is smaller than the temperature contribute to
the qubit dephasing. For a random telegraph process, $\xi(t)$, in
which a switching between the values $\pm 1/2$ takes place at random
times, the probability to make $n$ transition in a certain amount of
time $\tau$ is Poisson distributed. As a consequence, the
time correlation function
$\langle\xi(t)\xi(0)\rangle=e^{-|\gamma|t}/4$ decays exponentially,
with $|\gamma|$ given by the rate of the transition between the to
states of the fluctuator. The contribution to the noise spectrum of
a random telegraph process, given by the Fourier transform of the
correlation function, is a Lorentzian,
$\gamma/4\pi(\omega^2+\gamma^2)$. The effects of many uncorrelated
fluctuators coupled to the qubit via constants $v_i$ simply add. 
Considering a large number of effective spin-fluctuators and assuming
no correlations between couplings $v_i$ and switching rates
$\gamma_i$, the noise spectral function is give by
\begin{equation}
S(\omega)=\frac{\langle v^2\rangle}{4\pi}
\int\frac{\gamma{\cal P}_{\gamma}(\gamma)}{\gamma^2+\omega^2}
d\gamma.
\end{equation}    
The distribution ${\cal P}_{\gamma}(\gamma)$ depends on the details of
the interaction between the fluctuators and the qubit. Following
\cite{Galperin-PRB07,Galperin-04}, ${\cal P}_{\gamma}(\gamma)\sim
(A/\gamma)\Theta(\gamma-\gamma_0)$, where $\gamma_0$ represent the
maximal relaxation rate for fluctuators with a given energy splitting
$E$, and $A$ gives the amplitude of the $1/f$ noise. It follows for
the noise spectral function
\begin{equation}\label{Eq:1overfSpectralF}
S(\omega)=\frac{A}{\omega}\times\left\{\begin{array}{cc}
1 & , ~\omega\ll \gamma_0\\
2\gamma_0/\pi\omega & ,~\omega\gg\gamma_0
\end{array}\right..
\end{equation} 
The spin-fluctuator model reproduces $1/f$ noise power spectrum for
$\omega\ll\gamma_0$, but it predicts a crossover from a $\omega^{-1}$
to $\omega^{-2}$ behavior, consequence of the assumption of a maximal
switching rate $\gamma_0$. Expressing the fluctuation of the magnetic
flux as sum of the contributions of the independent fluctuators
$\Phi(t)=\sum_ib_i\xi_i(t)$, in a Gaussian approximation, substitution 
of Eq.~(\ref{Eq:1overfSpectralF}) in
Eq.~(\ref{Eq:Gaussiandecay1overf}) 
leads to
\begin{equation}
{\cal K}_g=-\ln\rho=\langle v^2_{\Phi}b^2\rangle At^2
\times\left\{\begin{array}{cc}
\gamma_0t/6 & , ~\gamma_0t\ll 1,\\
\ln 2 & ,~\gamma_0t\gg 1.
\end{array}\right.
\end{equation}
Therefore the echo signal is expected to decay in a Gaussian way at
long times $t\gg\gamma_0^{-1}$, while at short times
$t\ll\gamma_0^{-1}$ a faster decay is expected, ${\cal K}_g\propto
t^3$.   

Using the method of stochastic differential equations (see
\cite{Galperin-PRB07} and references therein), an
estimate for the non-Gaussian case provides for the logarithm of the
echo signal ${\cal K}_{\rm sf}$ \cite{Galperin-PRB07}
\begin{equation}
{\cal K}_{\rm sf}\approx
\left\{\begin{array}{cc}
\gamma_0A\bar{v}^2t^3/6 & ,  ~t\ll\gamma_0^{-1}\\
\ln 2A\bar{v}^2t^2 & ,~\gamma_0^{-1}\ll t\ll\bar{v}^{-1}\\
\alpha\bar{v}At & ,~\bar{v}^{-1}\ll t,
\end{array}\right.
\end{equation}
where $\bar{v}$ is the center of a sharp peaked distribution of the
couplings between the fluctuators and the qubit, and $\alpha\approx
6$. A new decaying behavior arise in the case of large time
$t\gg\bar{v}^{-1}$, that drastically differs from the predictions of
Gaussian statistically distributed magnetic flux fluctuations. The
reason laying in the fact at short time only fast ``fast'' fluctuators
contribute to the dephasing, giving rise, for $v\ll\gamma$, to
Gaussian decay, while at long time non-Gaussian behavior appears.

\subsubsection{$1/f$ noise in superconducting phase qubit}

\begin{figure}
 \begin{center}
 \includegraphics[width=7cm]{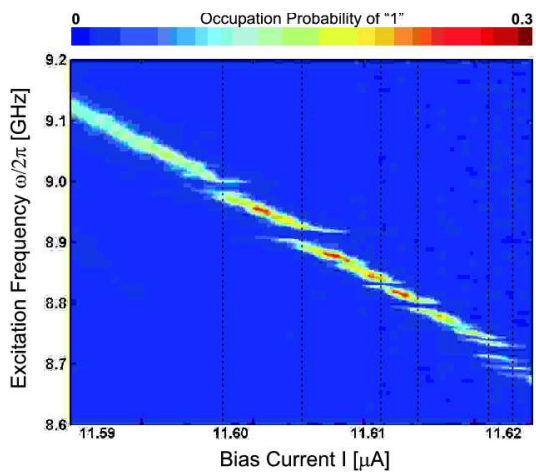}
    \caption{Measured probability of state ``1'' versus microwave
      excitation frequency $\omega/2\pi$ and bias current $I$ for a
      fixed microwave power for a phase qubit, obtained from
      Ref.~\cite{Simmonds-PRL04}. Dotted vertical line indicate
      spurious resonators. 
      (With permission from the Authors).
    \label{Fig10}}
 \end{center}
\end{figure}
In this section we review the latest results on decoherence of
superconducting phase qubits \cite{Ioffe-Nature99,Martinis-PRL02}.
Recent experiments \cite{Simmonds-PRL04,Martinis-PRL05} have
pointed out that a dominant source of decoherence for the phase qubit
is represented by low frequency $1/f$ noise, that is believed to arise
from two level systems (TLS) in the insulating barrier of the tunnel
junction as well as in the dielectric material surrounding the
circuit. In Fig.~\ref{Fig10} a measurement of the transition
frequency $\omega_{01}$ between the two qubit state, as a function of
the bias current $I$ and the microwave excitation frequency
$\omega/2\pi$, shows a qubit line in which a number of spurious
resonators appear, characteristic of energy-level repulsion predicted
for coupled two-state systems. Near the resonators, the Rabi
oscillations show beating, loss and recovery of the
oscillations with time, and rapid decrease of coherence amplitude. 
The beating behavior is consistent with the interaction of a qubit with
another two-level system, but not with harmonic oscillator modes in
the read-out SQUID. Moreover, each qubit has its own set of resonator
frequencies and strength, indicating that the TLS have a microscopic
origin. 

A new method to measure $1/f$ noise in Josephson junction qubits has
been recently presented in \cite{Bialczak-PRL07}. It uses the resonant
response of the qubit to directly measure the spectrum of the
low-frequency noise, and allows to distinguish between flux and
critical-current fluctuations by comparison of the noise taken at
positive and negative bias. Remarkably it can yield low-frequency
spectra below 1 Hz. Dephasing is produced by low-frequency
fluctuations in the qubit energy, which in this study are believed to
arise from magnetic flux noise in the qubit loop, with a spectral
density that scales inversely with frequency, $1/f$.  It turns out
that the is flux-like noise predominates over critical-current noise. 

The possibility that flux noise is due to TLS defects in the native
oxides of the superconductive film, as proposed in \cite{Koch-PRL07},
is examined in Ref.~\cite{Bialczak-PRL07}. Following
\cite{Shnirman-PRL05} for a standard TLS model \cite{Hunklinger81} a
theoretical estimation of the flux noise spectral density, for a
realistic geometry of the circuit loop, 
gives $S_{\Phi}(1~{\rm Hz})\approx10^{-3}(\mu\Phi_0)^2/{\rm Hz}$,
about 4 orders of magnitude smaller than the measured flux noise. This
estimate is based on the assumption that  TLS fluctuations randomizes
the defect magnetic moment; assumption highly questionable because TLS
defects in typical oxides are not considered to be magnetic.

\end{document}